# Micro/nanomaterials for improving solar still and solar evaporation - A review

Guilong Peng[1,#], Swellam W. Sharshir[1,2,3, #], Yunpeng Wang[1], Meng An[4], A.E. Kabeel[5], Jianfeng Zang[2], Lifa Zhang[6], Nuo Yang[1*]

1 State Key Laboratory of Coal Combustion, School of Energy and Power Engineering, Huazhong University of Science and Technology, Wuhan 430074, China

2 School of Optical and Electronic Information, Huazhong University of Science and Technology, Wuhan 430074, China

3 Mechanical Engineering Department, Faculty of Engineering, Kafrelsheikh University, Kafrelsheikh, Egypt

4 College of Mechanical and Electrical Engineering, Shaanxi University of Science and Technology, Xi'an, 710021, China

5 Mechanical Power Engineering Department, Faculty of Engineering, Tanta University, Tanta, Egypt

6 Center for Quantum Transport and Thermal Energy Science, School of Physics and Technology, Nanjing Normal University, Nanjing, 210023, China

[#]Guilong Peng and Swellam W. Sharshir contribute equally on this work.

*Corresponding email: nuo@hust.edu.cn

## Abstract

In last decades, solar stills, as one of the solar desalination technologies, have been well studied in terms of their productivity, efficiency and economics. Recently, to overcome the bottleneck of traditional solar still, improving solar still by optimizing the solar evaporation process based on micro/nanomaterials have been proposed as a promising strategy. In this review, the recent development for achieving high-performance of solar still and solar evaporation are discussed, including materials as well as system configurations. Meanwhile, machine learning was used to analyze the importance of different factors on solar evaporation, where thermal design was founded to be the most significant parameter that contributes in high-efficiency solar evaporation. Moreover, several important points for the further investigations of solar still and solar evaporation were also discussed, including the temperature of the air-water interface, salt rejecting and durability, the effect of solid-liquid interaction on water phase change.

**Keywords：** micro/nanomaterials; micro/nanoparticles; solar still; solar evaporation; desalination; phase change.

# 1. Introduction

Safe freshwater is essential for urban expansion, human civilization, and industrial development. Billions of people are suffering freshwater scarcity while seawater covers 70% of the earth in area. Given the both of environment pollution and water scarcity, it is desirable to develop eco-friendly technology for seawater desalination [1]. However, due to population growth and industrial development, human demand for freshwater resources has risen rapidly day after day. Nowadays, the problem of drinking water is one of the major problems that both developed and developing countries are facing [2]. Most health problems are caused by the scarcity of clean drinking water [3]. In recent decades, the lack of rainfall all over the world has led to an increase in the salinity of water bodies. Environmental pollution further exacerbates the scarcity of clean freshwater [4].

One of most important ways to address the freshwater scarcity is seawater desalination. Solar desalination, which uses renewable solar energy, is considered as a promising desalination technology to provide clean water with no or minimal environmental impacts. Compared to high-grade energy such as fuel and electric energy, solar energy does not depend on the long-range transportation due to worldwide distributed. At the same time, solar energy is eco-friendly energy, which is in agreement with the environmental protection policies all over the world. Therefore, solar desalination is getting more and more interest [5].

As a convenient solar thermal desalination technology, solar still is very useful in some places where insufficient electrical power and complicated desalination plants (such as RO plants) are not available [6]. For instance, poor, remote coastal area and emergency water supply in outdoor. The system and working principle of solar still are very simple compared to other desalination technologies, which make it no need skilled labor for operation and maintenance [7]. Meanwhile, the efficiency of the photo-thermal process is much higher than the photo-electric process, which gives solar still a potential to outperform electric based desalination process. However, the application of solar still is limited by its low productivity. Investigation of suitable materials and system designs might make photo-thermal process based solar still system develop to a new stage and contribute more in desalination field.

During past decades, many traditional ways have been developed to improve the efficiency of solar still, including using cotton cloth [8], sponge [9], charcoal [10] and so forth to improve the solar absorption. External area and power source, such as collectors [11], condenser[12] or reflectors [13], were also used and broadly studied. However, the energy efficiencies of traditional solar stills remain low (about 30-45%), due to the inefficient evaporation and condensation process, as well as the large heat dissipation to the ambient of system [14].

Recently, micro/nanotechnology shows a great potential in improving solar desalination and solar evaporation, due to high thermal conductivity, large surface area and high solar absorptivity of micro/nanomaterials [15]. Various nanostructured solar absorber materials, such as plasmonic metals [16], carbon-based materials [17], polymers [18] and semiconductors [19] with efficient photothermal conversion capabilities have been studied. Several effective concepts have been proposed to fully take the advantages of nanotechnology. For example, nanobubble formation [20] and heat localization [21], which direct the design of volumetric evaporation system and surface heating evaporation system, respectively [5, 22].

In this paper, the state-of-the-art developments in micro/nanotechnology for improving solar still and solar evaporation is reviewed. Various factors, such as material form, materials types, and thermal design and so forth, are compared and discussed in details. To evaluate the importance of different factors, a dataset of reported results is built and the machine learning method is carried out. The important gaps between research results of solar evaporation and solar still are also highlighted and discussed, which points out the future directions of this field.

# 2. Solar still by micro/nanoparticles

## 2.1 Passive solar still

The schematic diagram of how to use suspended nanoparticles or nanofluids in a basic solar desalination system, i.e. solar still (SS), is shown in Fig. 1a and 1b. It consists of a well-insulated container where brackish/saline water with nanoparticles is collected at its base. The upper surface of the container is covered by a tightly sealed glazier cover to minimize vapor leakage. Insolation entering the still through the glass is absorbed by nanofluids, hence the nanofluid is heated up. Nanoparticles absorb a large amount of solar radiation and discharge the heat to a base fluid (water) and the water molecules begin to evaporate faster. Vapor molecules are carried up from the surface of water to the air inside the still by natural convection and the air becomes saturated by vapors. When this saturated air strikes the cool inner glass surface, condensation of vapor molecules began to occur. This condensate water slides down due to gravity and accumulates outside the still as freshwater.

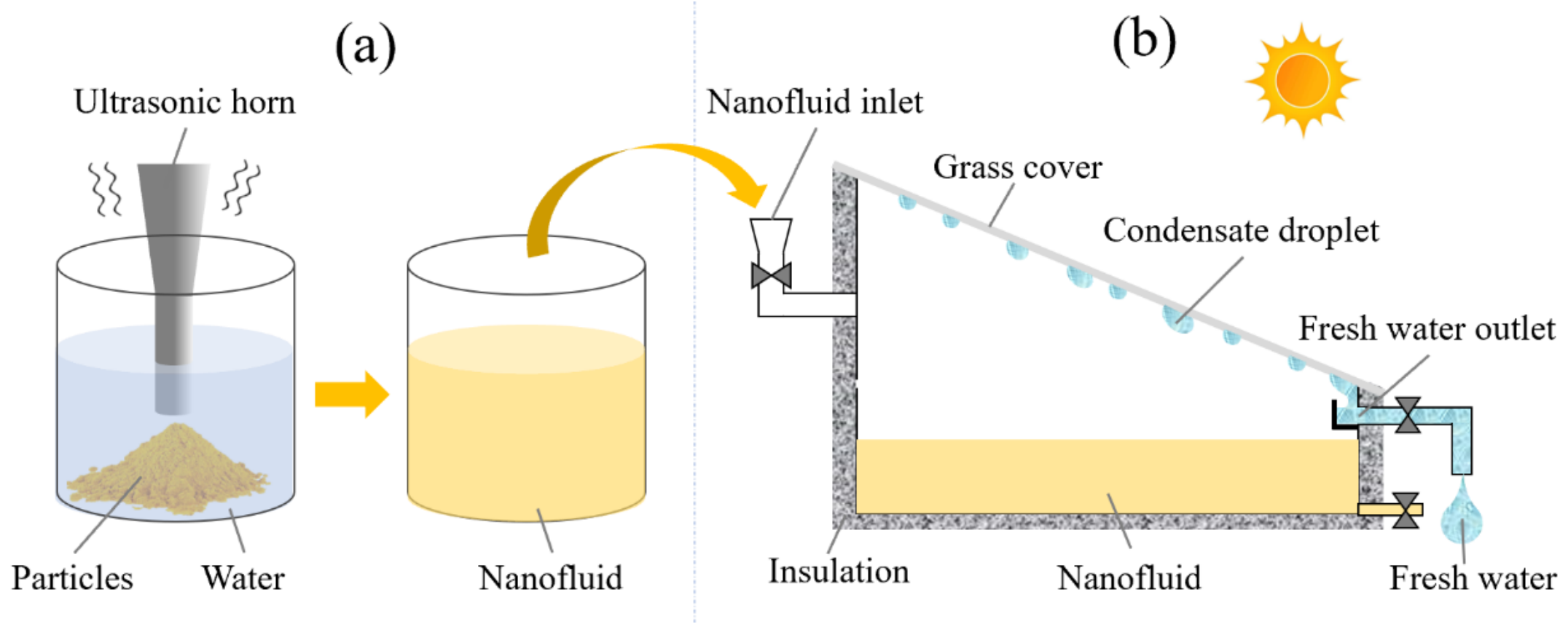

Fig. 1 (a) Mixing nanoparticles with base water and (b) Working conventional solar still with nanofluid.

As illustrated in Fig. 1b, passive solar still indicates the most basic solar still, which has no extra power input and additional equipment. In this section, the modifications done in the passive solar still with nanoparticles are summarized. Oxide nanoparticles based nanofluids are the most common type used in solar still, due to the excellent durability and low cost. $Al_2O_3$ was found to be an effective nanoparticle among various oxide nanoparticles in passive solar still. One of the reasons might be that higher thermal conductivity of nanofluid gives higher efficiency and productivity under sunny days in April at Kovilpatti (9° 11′ N, 77° 52′ E) Tamil Nadu, India, $Al_2O_3$ nanofluid gave 29.95% enhancement, pursued by 18.63% for $SnO_2$ and 12.67% for ZnO. Thermal conductivity of $Al_2O_3$, $SnO_2$, and ZnO nanofluids are 0.6355

W/($m^2$·K), 0.6215 W/($m^2$·K), 0.6105 W/($m^2$·K), respectively [23]. The trend of thermal conductivity agrees well with the enhancement. The experiments of CuO, $TiO_2$ and $Al_2O_3$ under climatic conditions of New Delhi for the month of March also give the same conclusion[24]. The test has been carried out at various concentrations (0.2, 0.25 and 0.3%wt) in passive double slope SS to obtain the best concentration (0.25%) of the three nanoparticles. The best energy efficiency was gained for nanofluid (CuO 43.81%, $TiO_2$ 46.10%; and $Al_2O_3$ 50.34%;) in compared with the pure water (37.78%). The results also agree with the trend thermal conductivity which are 0.6901 W/($m^2$·K), 0.7261 W/($m^2$·K) and 0.7863 W/($m^2$·K), for CuO, $TiO_2$ and $Al_2O_3$, respectively. The experimental results agree with the theoretical analysis, which further proves the importance of thermal property of nanofluid in solar still.

Nevertheless, the concentration of nanofluid in the above-mentioned works are relatively low (0.05-0.3%wt), hence the enhancement is not very high. Enhance the concentration might further enhance the productivity of solar still [3]. However, given the white color of $Al_2O_3$, $TiO_2$, ZnO and $SnO_2$ nanoparticles, the high reflectance of these nanoparticles under high concentration might decrease the productivity. Therefore, nanoparticles with black color might be a better choice for a higher concentration of nanofluid in solar still, such as CuO and carbon-based nanoparticles. Sharshir et al. investigated the effect of graphite micro-flakes (GMF) and copper oxide (CuO) with glass cover cooling on the SS performance on September in Wuhan, China. The GMF and CuO with various weight concentrations ranged from 0.125% to 2% was studied. Various brine depths ranged from 0.25 to 5 cm., and various glass cover cooling flow rates ranged from 1 to 12 kg/h were examined to achieve the best performance. The obtained results using CuO with glass cooling enhanced the still productivity by approximately 44.91% as well as using GMF with glass cooling improved the productivity by 53.95%, compared with the conventional one. When using the CuO and the water exit from glass cover cooling as feed water to the still the freshwater is enhanced by approximately 47.80%. Eventually, the daily efficiencies of the modified SSs using GMF and CuO microparticles with glass cooling are 49% and 46%, respectively [25]. Furthermore, when GMF nanofluid is combined with glass film cooling and phase change material the daily productivity is enhanced by 73.80%, as compared with the conventional SS [26].

## 2.2 Active solar still

The active solar still is similar to the passive solar still but integrated with the external energy dissipation devices such as external condenser and fan or collectors and pump as illustrated in Fig. 2. In this section, the modifications done in the active solar still with micro/nanoparticles are summarized. Due to the enhanced the heat and mass transfer rate, which is a benefit to the evaporation and condensation process, active solar still is found to be able to take better advantage of nanofluids than passive solar still.

External condenser is one of the most popular and effective components used in active SS for enhancing the condensation rate. By combining nanofluid and external condenser, both high evaporation rate and condensation rate can be achieved. Kabeel et al. studied the effect of $Al_2O_3$ and $Cu_2O$ on the performance of SS with and without external condenser. The performance was examined at various concentrations without and with providing vacuum. The large improvement in output freshwater productivity was gained with using $Cu_2O$ nanoparticles at the optimal concentration of 0.2% together with the vacuum fan as an external condenser. The experimental results showed that using $Al_2O_3$ nonmaterial increased the freshwater output approximately by 88.97% and 125.0% without and with electric fan (external condenser), respectively, as compared with that of conventional stills. $Cu_2O$ nanoparticles improve freshwater output by about 93.87% and 133.64% without and with electric fan (external condenser), respectively [27]. Omara et al. investigated experimentally a hybrid distillation unit including corrugated plate absorbers, wick material, reflectors, external condenser and nanoparticles ($Cu_2O$ and $Al_2O_3$). The productivity of modified SS was enhanced by approximately 285.10% and 254.88% for $Cu_2O$ and $Al_2O_3$, respectively, at 1cm brine depth and 1.97%wt. of concentration [28]. Clearly, active SSs with nanofluids show a much better performance than traditional passive SSs.

However, the cost might increase due to the need of extra devices and energy.

The nanofluid-based solar collector is another popular component for improving active solar still. Compared to directly use nanoparticles in the basin of solar still, nanoparticles in solar collector can be more stable due to the stir of pumps. The loss of materials can be also avoided when the flow cycle is closed as illustrated in Fig. 2. Nevertheless, the productivity enhancement of using nanofluid-based solar collector is not very significant according to the reported results. Mahian et al. conducted investigation to improve SSs by using two types of nanofluids ($SiO_2$ and Cu /water nanofluid) with two flat plate collectors joint in series, as illustrated in Fig. 2. It shows that the productivity of SS improved by only about 1% when using $SiO_2$ nanofluid at 4% volume fraction and nanofluid temperature at 70 ºC, although the convection heat transfer coefficient is enhanced by about 15.4% [29]. Sahota et al. also investigated the active SS integrated with water collectors and heat exchanger. The freshwater output of using CuO, $TiO_2$, and $Al_2O_3$ was enhanced by about 31.49%, 7.26% and 26.4%, respectively, compared with that of pure water [30].

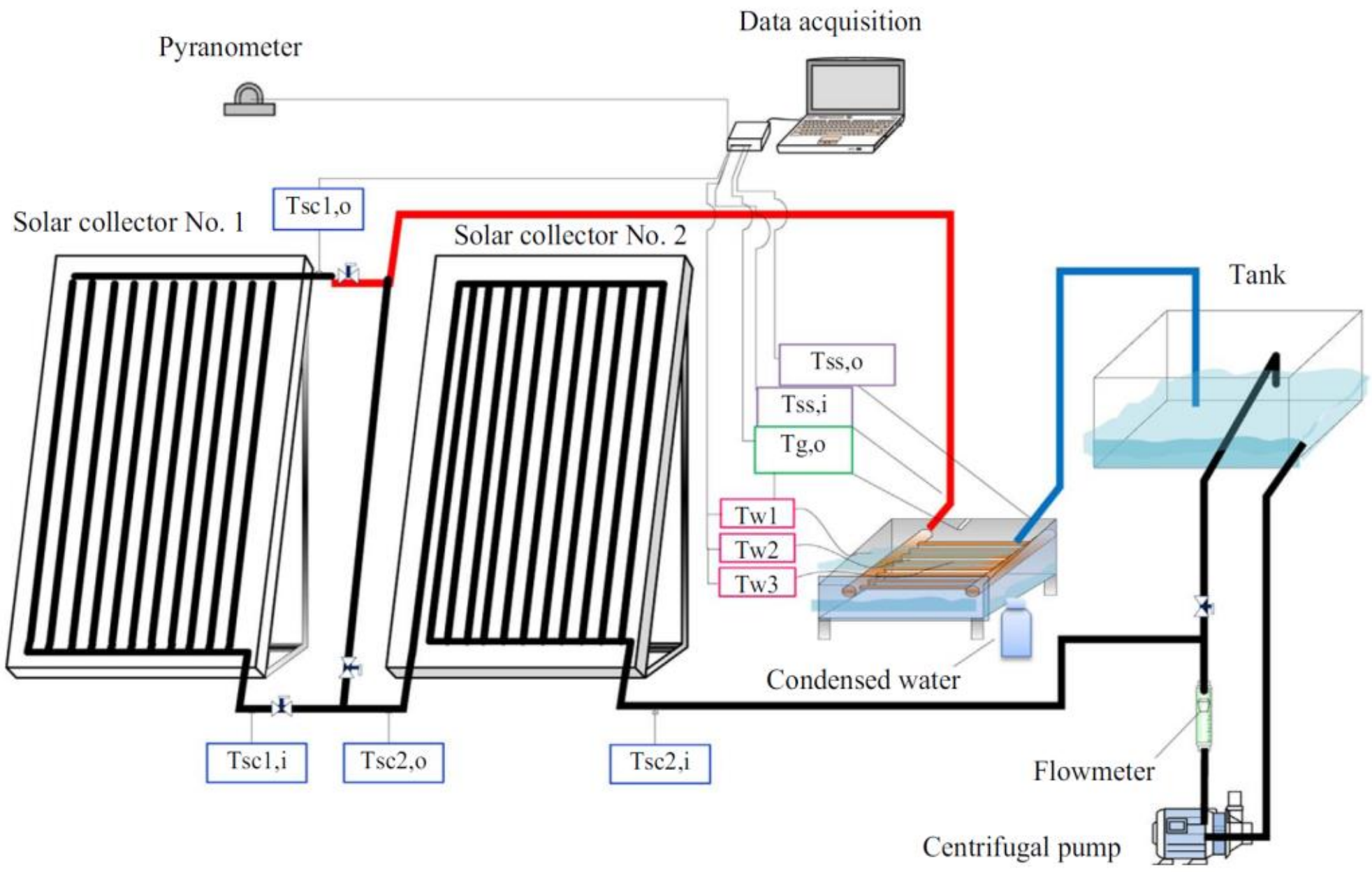

Fig. 2 Diagrammatic of the experimental setup of solar still with the solar collector and nanofluid [29]. Copyright 2017 Elsevier.

## 2.3 Challenges of using nanofluids in solar still

In solar stills, especially in active solar stills, metallic surfaces are important to improve the heat transfer process. However, nanofluid might enhance the erosion and corrosion on a metallic surface by both chemical and physical ways [31]. When the fluid's characteristics fall in the typical chemical corrosion range, even for a limited time interval, a marked and fast consumption of the metallic surface will be observed [32]. Meanwhile, nanofluid also shows erosion due to the collision between the metallic surface and particles in the bended pipes [33]. Therefore, to avoid the erosion and corrosion, the nanofluid should be preliminarily tested before its adoption in desalination system and be assessed on possible negative interactions with components [31]. It is also effective to take advantage of nanofluid itself to decrease erosion and corrosion, such as promoting the formation of a compact protective film on the metallic surface by nanoparticles [34]. Proper system design and maintenance are also necessary to decrease effects of erosion and corrosion [35].

Besides erosion and corrosion phenomena, stability of nanofluid and pressure drop are another two problems when using nanofluids for solar desalination. Long term stability is still one of the major challenges for nanofluids and requires more researches. Poor stability results in aggregation and settlement of particles as well as chemical dissolutions, which lead to the failure of nanofluid [36]. For passive device, where there is no pump to circulate and stir the nanofluids, high agglomeration of nanoparticles will occur, specially at height temperature gradients [37]. Pressure drop and pumping power problems will appear in active solar still when using nanoparticles. The increase of nanofluid concentration results in an increase in pressure drop under turbulent regime [38]. The increased pressure drop will inevitably increase the operation cost of system.

## 2.4 Section summary

Nanofluids have great potential to enhance the energy efficiency and heat transfer in solar still system, which increases the freshwater output of solar still system. Many kinds of nanoparticles are studied in the previous works, such as $Al_2O_3$, ZnO, $SnO_2$, CuO, $TiO_2$ and graphite particles (Table 1). The enhancement of productivity is due to the several merits compared to base fluid (i.e. water), such as high thermal conductivity and high solar absorptivity. However, the stability of nanofluid, erosion and corrosion by nanofluid, as well as pressure drop of pump by nanofluid are three important challenges in this field. Meanwhile, oxide nanoparticles such as $Al_2O_3$, ZnO and $TiO_2$ are white, which should have low solar absorptivity thus to the disadvantages of enhancement in productivity. Therefore, more detailed and fundamental studies should be explored to further understand the reported enhancement by these white nanoparticles.

**Table 1.** Summary of solar stills with different types of micro/nanoparticles for enhancing the production

| Types of SS | Types of particles | Size | Concentration, % | Modifications | Enhancement, % | Criticism and disadvantages |
|---|---|---|---|---|---|---|
| Passive solar stills | ZnO [23]<br>$Al_2O_3$<br>$SnO_2$ | 9.3-16 nm<br>240-395 nm<br>114-115 nm | 0.1 | - | 12.67<br>29.95<br>18.63 | missing efficiency |
| | Graphite [25a]<br>CuO | 1.3 µm<br>1 µm | 1 | Glass cooling | 57.60<br>47.80 | missing cost |
| | Graphite [26] | 1.3 µm | 0.5 | Glass cooling and PCM | 73.8 | missing efficiency |
| | $Al_2O_3$ [27b]<br>$Cu_2O$ | 10-14 nm | 0.2 | External condenser | 88.97<br>93.87 | missing efficiency |
| | $Al_2O_3$ [27a] | 10-14 nm | 0.2 | External condenser | 76 | missing efficiency |
| | $Al_2O_3$ [3] | 20 nm | 0.12 | - | 12.2 | missing cost |
| | CuO [24]<br>$TiO_2$<br>$Al_2O_3$ | 20 nm | 0.25 | - | 43.81<br>46.10<br>50.34 | missing cost |
| Active solar stills | $Al_2O_3$ [27b]<br>$Cu_2O$ | 10-14 nm | 0.2 | External condenser<br>With fan | 125.0<br>133.64 | missing efficiency |
| | $Al_2O_3$ [27a] | 10-14 nm | 0.2 | External condenser<br>With fan | 116 | missing efficiency |
| | $Al_2O_3$ [28]<br>$Cu_2O$ | 10-14 nm | 1.97 | Wick absorbers， internal reflectors and external condenser. | 285.10<br>254.88 | missing efficiency |
| | CuO [30]<br>$TiO_2$<br>$Al_2O_3$ | 20 nm | 0.25 | Flat plate collectors | 31.49<br>7.26<br>26.4 | missing efficiency |
| | $SiO_2$ [29]<br>Cu | 7 nm | 4 | Flat plate collector and heat exchanger | 0.66<br>9.86 | missing cost |

## 3. Solar evaporation by micro/nanomaterials

Recently, solar evaporation draws much more attention compared to the whole solar still system, particularly during the last five years. Much effort has been done to improve solar evaporation by nanomaterials. The solar evaporation system can be divided into two categories: a volumetric system and interface system, which contains nanofluid and floatable porous materials, respectively (Fig. 3). Porous materials are foams and membranes, which have micro/nanostructures.

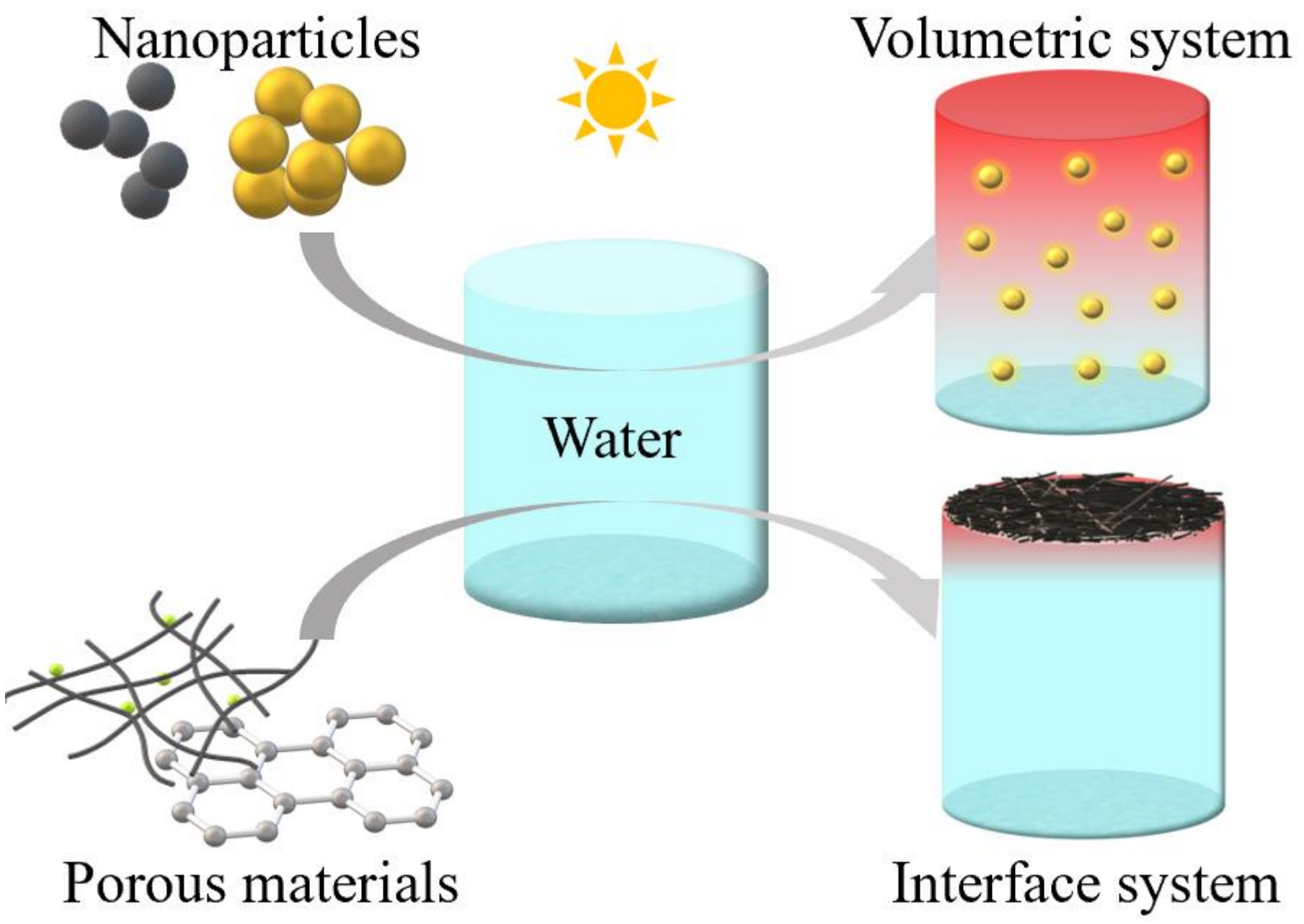

**Fig. 3** Schematic diagram of micro/nanomaterials in solar evaporation system.

### 3.1 Volumetric system by nanofluids

In a volumetric system, solar irradiation is mostly absorbed by suspended nanoparticles. Water is heated by nanoparticles and then evaporates. Metallic nanoparticles, carbon-based nanoparticles and are the two main kinds of nanoparticles for improving solar evaporation during the past decade. Metallic nanoparticles convert solar irradiation to heat mostly based on plasmonic effect, i.e. photons induced electronic resonance in metallic nanoparticles which generate heat due to electron-phonon scattering. Metallic nanoparticles strongly absorb light when the optical frequency matches the resonance frequency of electron in nanoparticles[5]. Meanwhile, the plasmonic effect in nanoparticles can be designed and adjusted by changing the shape, size, location, surface chemistry of particles, and so on [39]. Therefore, tuning the optical property of metallic nanoparticles becomes a hot topic during the last several years.

In the beginning, metal-based nanofluids arise much interest in evaporation field due to the laser-induced generation of vapor bubbles [40]. When the laser intensity overs a typical threshold, the water around the particles reaches a high temperature, resulting in explosive evaporation. Thus, a bubble around the nanoparticles will be formed [41]. The nanoparticle-generated bubbles may temporally and spatially localize laser-induced thermal field and prevent residual heating of the bulk media, which indicates a high energy efficiency, due to a lower bulk temperature and a less heat loss [20, 42].

Inspired by laser-induced nanobubbles generation, metal-based nanofluids were further studied under concentrated solar irradiation to show its potential for solar evaporation. By using golden nanofluids, Halas et al. found that 80% of the absorbed sunlight was converted into water vapor and only 20% of the absorbed light energy was converted into heating of the surrounding liquid [20, 43]. To understand the mechanism, subsequent researches were carried out. The results show that the high-efficiency evaporation is caused by the collective effect mediated by multiple light scattering from dispersed nanoparticles. Randomly positioned nanoparticles that both scatter and absorb light can concentrate light energy into mesoscale volumes near the illuminated surface of the liquid. The resulting light absorption creates intense localized heating and efficient vaporization of the surrounding liquid [44]. A similar conclusion was obtained by Jin et al. [45]. It means that the nanobubble doesn't exist in the solar irradiation-based system, which is different from the laser-based system.

To further improve the energy efficiency of metal-based nanofluid system, many efforts have been devoted to optimize the optical property of the nanofluid, such as controlling morphology [46], concentration [47], compounds of particles [48]and diameters [49]. However, the required power density for efficiently using metal-based nanofluid is always hundreds of suns, thus high equipment cost is required for the solar concentration. Meanwhile, the material cost of metal nanoparticles is quite high. Therefore, the potential of using metal-based nanofluid in industrial solar evaporation process is questionable.

Different from metal-based nanoparticles, carbon-based nanoparticles have much lower materials cost and higher solar absorptivity. The cost of producing 1 g/s vapor by gold nanoparticles is found to be ~300 folds higher than that produced by carbon black nanoparticles [50]. Carbon-based materials absorb solar energy mainly by thermal vibrations, hence a high and broadband absorptivity can be achieved under low solar intensity [5]. Therefore, some researchers turned to carbon-based nanoparticles or nanocomposites which shows a high efficiency under a relatively low solar concentration [51]. For example, graphene-silver nanoparticle composites exhibit a high efficiency as 80% under around 60 suns [52], and graphene oxide-gold nanoparticle composites have an efficiency up to 59.2% under 16.77 suns [53]. The pristine carbon nanoparticles such as carbon black, graphite and carbon nanotube have an efficiency up to 69% under only 10 suns [22]. Thereby, carbon-based nanoparticles seem to be a more suitable material for solar evaporation compared to metallic nanoparticles.

## 3.2 Interface system based on floatable porous materials

Compared to nanofluid, an essential benefit of floatable porous materials is creating solar heating at air-water interface. Key components are solar absorber, floating evaporation structure and porous materials. A solar absorber that can efficiently absorb and convert the solar radiation into heat. While allowing the vapor to permeate through the front face, a floating evaporation structure that can simultaneously maximize the evaporation rate and supply liquid to the heated region [54]. Porous materials, such as foams and films, concentrate thermal energy and fluid flow were needed for phase change and minimizes dissipated energy [21]. As shown in Fig. 3, solar energy is absorbed on the top surface of porous material and creates a hot layer. The porous foam, which has low thermal conductivity, prevents the heat transferring from hot layer to bulk water. That is, heat is localized at the hot surface. To reach a high evaporation rate, the setup for heat localization needs to have four main characteristics: high absorption in the solar spectrum, low thermal conductivity to suppress heat dissipation, hydrophilic surfaces to leverage capillary forces and promote fluid flow, and interconnected pores for fluid flow [21]. Moreover, compared to nanofluid, floatable porous materials can be easily recycled. Hence, the maintenance cost is low.

### 3.2.1 Materials design

#### (i) Films

Due to the thin thickness (mostly micrometer scale), films show a potential to achieve high evaporation efficiency with very few amounts of materials. Paper-based film is one of most popular materials used in the air-water interface evaporation system, because it is cheap and scalable [55]. A common method to fabricate paper-based photo-thermal film is depositing nanoparticles on airlaid paper. For example, gold nanoparticles and graphene oxide nanoparticles [56] [57]. Due to the improved solar absorption by higher surface roughness of papers, paper-based AuNP films give a much higher evaporation rate compared to pure AuNP films. The energy efficiency of paper-based AuNP film is able to reach 77.8% under 4.5 suns [58].

Other kinds of porous films, such as silica membrane [19], noble metal membrane [59], aluminum oxide membrane [60], zeolite membrane [61], wood membrane [62], polymer and fibers membrane[63], are also well studied. Some films exhibit good performance with a thin thickness. The nitrogen doped 3D porous hydrophilic graphene membrane (thickness 35 μm) enables an efficiency reaches up to 80 % under one sun of solar irradiation [64]. The evaporation efficiency of MXene thin membranes with only several micrometers in thickness can reach up to 84% under one sun irradiation [65]. Based on alumina nonporous membrane (50 μm in thickness), Zhu's group developed several plasmon-enhanced solar desalination devices. Firstly, gold nanoparticles were used as plasmonic absorbers and found that the energy efficiency can reach up to over 90% under 4 suns. Later, to decrease the materials cost, gold nanoparticles were replaced by aluminum nanoparticles. The energy efficiency remains over 90% under 6 suns. [39, 66]. It is also found that films with nanometer thickness also show high efficiency. The ultrathin 2D porous photothermal film (120 nm in thickness) based on $MoS_2$ nanosheets and single-walled nanotube (SWNT) gives an energy efficiency reach up to 91.5% under 5 sun [67].

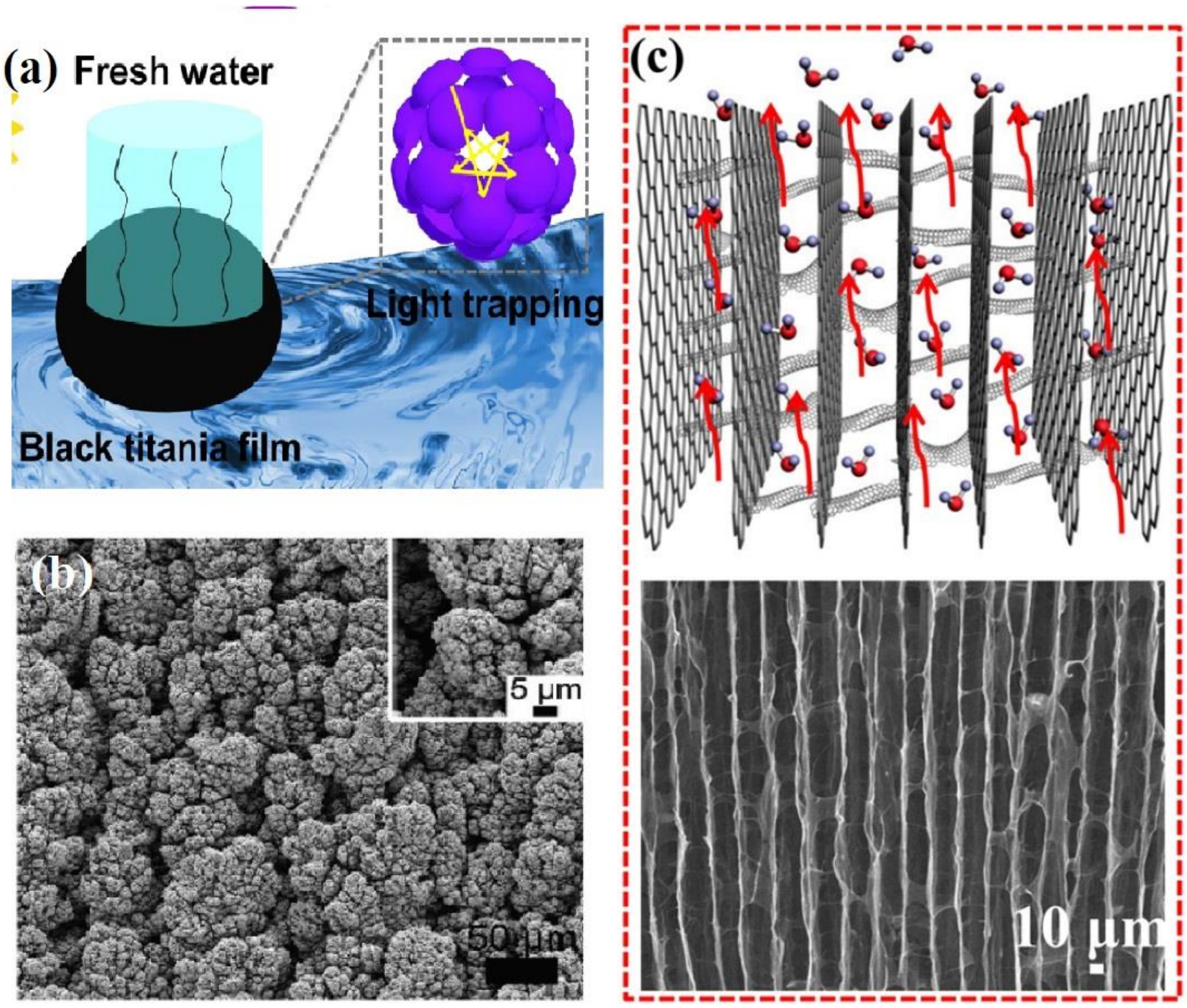

Fig. 4 Different nanostructure morphologies of membrane. (a) Nanocage of black Titania nanoparticles. (b) Cauliflower-shaped hierarchical surface nanostructure on a copper surface. (c) Vertically aligned graphene sheets membrane (VAGSM). (a) Reprinted with permission from Ref. [68]. Copyright (2016) American Chemical Society. (b) Reprinted with permission from Ref. [69]. Copyright (2016) The Royal Society of Chemistry. (c) Reprinted with permission from Ref. [70]. Copyright (2017) American Chemical Society.

Besides the use of different materials, the designing of micro/nanoscale membrane surface structures is also a popular strategy to enhance evaporation. The main purpose of structure designing is to enhance solar absorptivity. It is challenging to achieve high solar absorptivity in a very thin film for most materials, due to shortened light path. Therefore, to absorb as much light as possible within a limited thickness, the light trapping effect has been proposed in many works, which enhances residence time and length of light path [68]. For example, a black Titania film with unique nanocage structure on the surface can increase the energy efficiency by more than 30% under solar intensity of 1 $kW/m^2$, due to the dramatically enhanced absorptivity (Fig. 4a) [68]. Other special structures, such as cauliflower-shaped hierarchical surface (Fig. 4b), vertically aligned graphene sheets membrane (VAGSM) (Fig. 4c) were also proved very effective in improving solar evaporation. [69-70]

## (ii) Foams

Although films exhibit excellent performance, its thin thickness limited the heat localization effect; hence the heat loss to bulk water remains relatively high. Therefore, to further prevent the heat transfer between hot interface and bulk water, foams (centimeter scale in thickness) are a better candidate than films.

Aerogels are ideal materials for solar evaporation, due to their extremely low thermal conductivities and porous structures. One of the ways to fabricate suitable aerogels is integrating cellulose with metallic nanomaterials, such as dispersing gold nanorods into highly porous bacterial nanocellulose based aerogels [71]. Another way is integrating cellulose with carbon materials which have broader absorption band and more cost effective compared with metallic materials. The bilayer structure is a common method to use carbon materials in cellulose. A layer of the carbon material is on the top for solar absorption, while a layer of aerogel under is used for water transport, thermal insulation and supporting the carbon layer. The evaporation efficiency can reach up to 78% at 1 $kW/m^2$ solar irradiation, which is much higher than metallic nanomaterials based cellulose aerogel [72]. Fabricating carbon aerogel directly is also well studied. GO aerogel and graphene aerogel is the most popular carbon aerogel used for solar evaporation [73]. The evaporation efficiency is up to around 80%-90% under only 1 sun. [74].

Interestingly, hierarchically structured aerogels show a greater potential in enhancing solar evaporation compared to other aerogels. The hierarchical graphene foam (h-G foam) with continuous porosity grown via plasma-enhanced chemical vapor deposition, shows energy efficiency of solar evaporation up to 93.4% (Fig. 5a)[75]. The hierarchically nanostructured gels (HNG) evaporates water with a record high rate of 3.2 kg/($m^2$·h), and the energy efficiency reaches up to 94% at one sun irradiation[76]. This extremely high evaporation rate was 2.1 times that of the traditional limitation of evaporation rate, which is around 1.5 kg/($m^2$·h) under 1 sun [77]. Coincidentally, the hierarchical graphdiyne-based architecture also provides a high energy efficiency as 91% under 1 sun[78]. Those results indicate that the hierarchical structure may have a great impact on the solar evaporation process. However, the underlying mechanism remains to be uncovered.

Apart from the artificially synthesized aerogels showed above, natural biomaterial-based foam was also proved efficient in solar distillation, such as mushroom and wood (Fig. 5b and 5c) [79]. The cell walls in biomaterial form natural porous structure which is similar to that of aerogel, thus biomaterial also has a low thermal conductivity which is important for heat localization. Meanwhile, the cellulose in biomaterial is hydrophilic and provides strong capillary force for water replenishment to evaporation surface. Those characters of biomaterial and inspired a broad research of using woods in solar evaporation and desalination. However, although the solar thermal efficiency could be relatively high by using biomaterials, the durability of biomaterial is questionable, due to corrosivity of seawater to biomaterials.

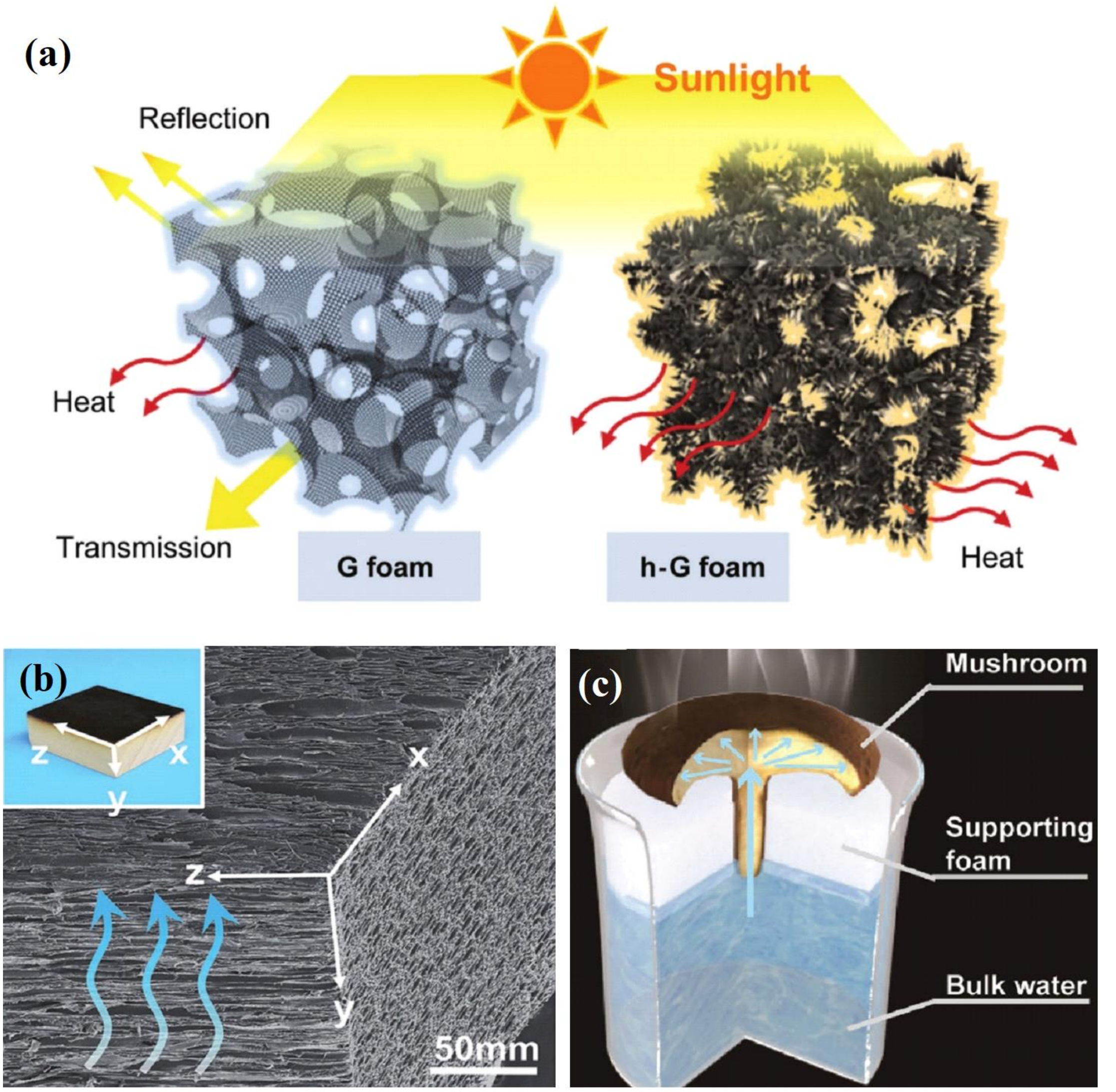

Fig. 5 (a) Schematic diagram of hierarchical graphene foam. G foam means graphene foam, h-G foam is hierarchical graphene foam [75]. (b) Surface carbonized wood with the tree-growth direction parallel to the water surface [80]. (c) Schematic of a mushroom-based solar steam-generation device [79c]. (a) Reprinted with permission from Ref. [75]. Copyright 2018 Wiley-VCH Verlag GmbH & Co. KGaA. (b) Reprinted with permission from Ref. [80]. Copyright 2018 Wiley-VCH Verlag GmbH & Co. KGaA. (c) Reprinted with permission from Ref. [79c]. Copyright 2017 Wiley-VCH Verlag GmbH & Co. KGaA.

One of strategies by using biomaterials in solar evaporation and desalination is coating surface of biomaterials by other materials, such as graphene oxide[81], carbon nano tubes [82], metal nanoparticles [83] and polymers [84]. Another more convenient and cost-effective strategy is carbonizing biomaterials directly [79b, 79c, 80]. Porous woods float on water spontaneously and function as insulation layer and water channel, while the black layer absorbs solar irradiation for water evaporation. Based on these principle, various biomaterials are studied, such as basswood [81], poplar, pine, and cocobolo wood [85], mushroom [79c], sugarcane [86] and leaf [87]. The most efficient biomaterials show an evaporation efficiency as high as 87% under one sun, which is comparable to artificial aerogels [84, 86].

### 3.2.2 Thermal design

Besides material design, thermal design of the system is also an effective way to improve the performance of solar evaporation. The thermal design of the evaporation system can be divided into three categories, 3D, 2D and 1D system, based on the dimension of water channel.

In 3D water channel systems, all the foam is penetrable, and water is transported through the bottom to the top surface of the foam for evaporation via the connected holes (Fig. 6a). Both aerogels and biomaterial-based systems, as discussed above belong to 3D water channel systems. Although the thermal conductivity of dry penetrable foam is low, the thermal conductivity increases obviously when the foam is filled with water, which weakens heat localization effect [21, 81].

Whereas in 2D water channel systems, a penetrable layer wraps over an impenetrable foam. The penetrable layer functions as a water channel and the impenetrable foam functions as an insulation layer. Because of reduced dimensionality of water channel, the heat dissipation through water will be decreased compared to 3D water channel systems. Li et al. firstly reported that with a 2D water channel, both efficient water supply and suppressed parasitic heat dissipation could be achieved simultaneously. The efficiency was reached to 80% by using graphene oxide film and polystyrene foam (Fig. 6b) [88]. Due to excellent thermal performances, the evaporation efficiency of the 2D water channel system can reach up to 88% by using cheap and scalable materials. However, the reduced dimensionality of the water channel might give rise to the problem of salt accumulation on the evaporation surface, when applied to seawater evaporation system[77]. In order to avoid salt accumulation, the molecular diffusion and water convection in the 2D water channels must be sufficient[89].

As for 1D water channel systems, there are two categories: bundle type system and truck type system (Fig. 6c and 6d). In the bundle type system, the water is transported through a bundle of 1D water channels, which is similar to the jellyfish (Fig. 6c). By using porous carbon black/graphene oxide (CB/GO) composite as the top solar absorption layer, aligned GO pillars as water channels, expanded polystyrene (EPS) as an impenetrable insulation layer. The assembled jellyfish-like evaporator can display a high energy efficiency of 87.5% under one-sun illumination [90]. On the other hand, truck type system contains only one 1D water channel, similar to the tree trunk [91]. Based on this concept, Liu et al. achieved 91.3% of energy efficiency under 1 sun by using airlaid paper as the truck and carbonized wood as solar absorber (Fig. 6d) [92]. Generally, both 2D and 1D water channel systems can reach a high energy efficiency with simpler material fabrication process, due to the minimized heat dissipation.

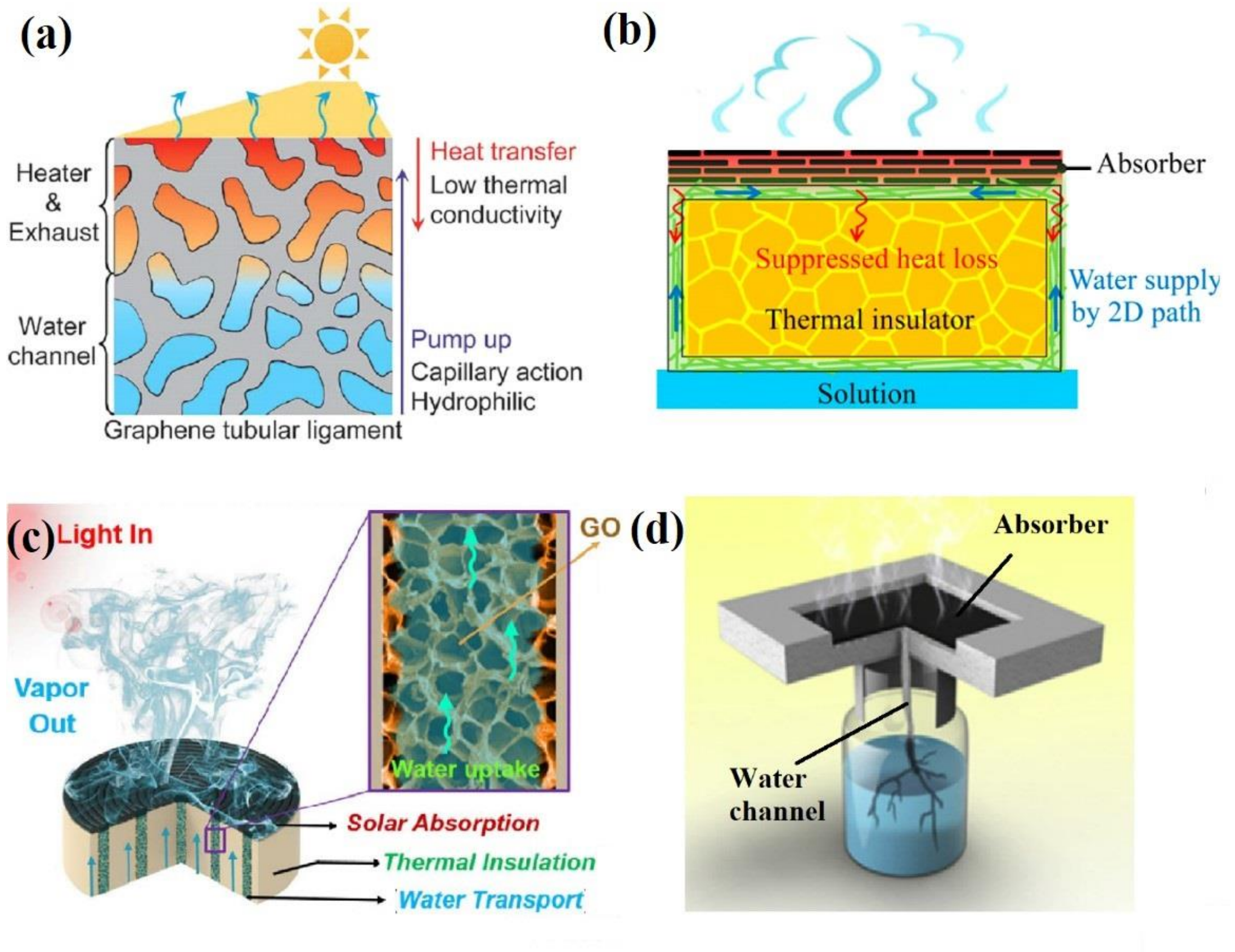

Fig. 6 Different thermal designs of solar evaporation system. (a) 3D water channel system, porous foam is full of connected holes which function as water channels during the evaporation process[64]. Copyright 2015 Wiley-VCH Verlag GmbH & Co. KGaA. (b) 2D water channel system, a penetrable layer wraps over impenetrable insulation foam. Water is transported along the wrapping layer[88] Copyright 2016, National Academy of Science.. (c) Root type 1D water channel system, water is transported along the root-like pillar array[90]. Copyright 2017, Elsevier (d) Trunk type 1D water channel system, the water channel is limited in the centre of impenetrable insulation foam[92]. Copyright 2018, Elsevier

## 3.3 Factors analyzing by machine learning

In the experiments of solar evaporation, the evaporation efficiency is affected by many factors, such as materials, thermal design, ambient temperature, and solar intensity, and so forth. Therefore, it is difficult to directly conclude the importance of different factors, and different works can’t be compared with each other. Herein, the machine learning algorithm of random forest (RF) is used to analyze the importance of various factors (i.e. descriptor in RF). RF has been widely applied in many scientific and engineering fields [93], the main step of RF is shown in Fig. 7. After model construction, the importance of certain descriptor can be calculated by intrinsic metric. Experimental data used in RF are collected from articles since 2014 (Table S2). The detail of method and dataset are listed in Supporting Information.

According to the results from RF, among all calculated factors, thermal design is the most important factor in solar evaporation, no matter how many descriptors are used in the calculation. The importance of thermal design is around 2 times higher than other descriptors (Fig. 8). This result shows that optimizing the heat transfer process in solar evaporation system is essential for enhancing the efficiency of solar evaporation. Solar absorptivity of materials is also relatively important, which is reasonable due to higher absorptivity enables more available energy for evaporation. However, most of the reported works have a very high absorptivity (>90%), hence its importance might be underestimated in the calculation. Meanwhile, due to the optimized thermal and material design, high efficiency can be obtained under low solar intensity as reported in many works (Table S2). Therefore, solar intensity is not important and similar to a random descriptor (i.e. a set of random data which should have no relationship to evaporation efficiency). The temperature of ambient ($T_{amb}$) and evaporation interface ($T_{interface}$) are also insignificant, which might due to the small difference of $T_{amb}$ and $T_{interface}$ between most of works as summarized in Table S2. The stable trend of importance of different factors in Fig. 8 proves that the calculated results are reasonable.

However, it should be noted that some papers do not provide information about all descriptors. Data of ambient temperature, the diameter of evaporation surface, absorptivity, and the temperature of evaporation interface are missing in some papers. Therefore, to obtain a more accurate result, it requires authors to provide complete information on experimental factors in their papers. On the other side, some other important factors, such as the material design, can’t be considered and calculated in the current stage, because the detailed properties of materials are not offered in most of papers, such as thermal conductivity, contact angle, specific area, porosity, characteristic size and chemistry properties and so forth.

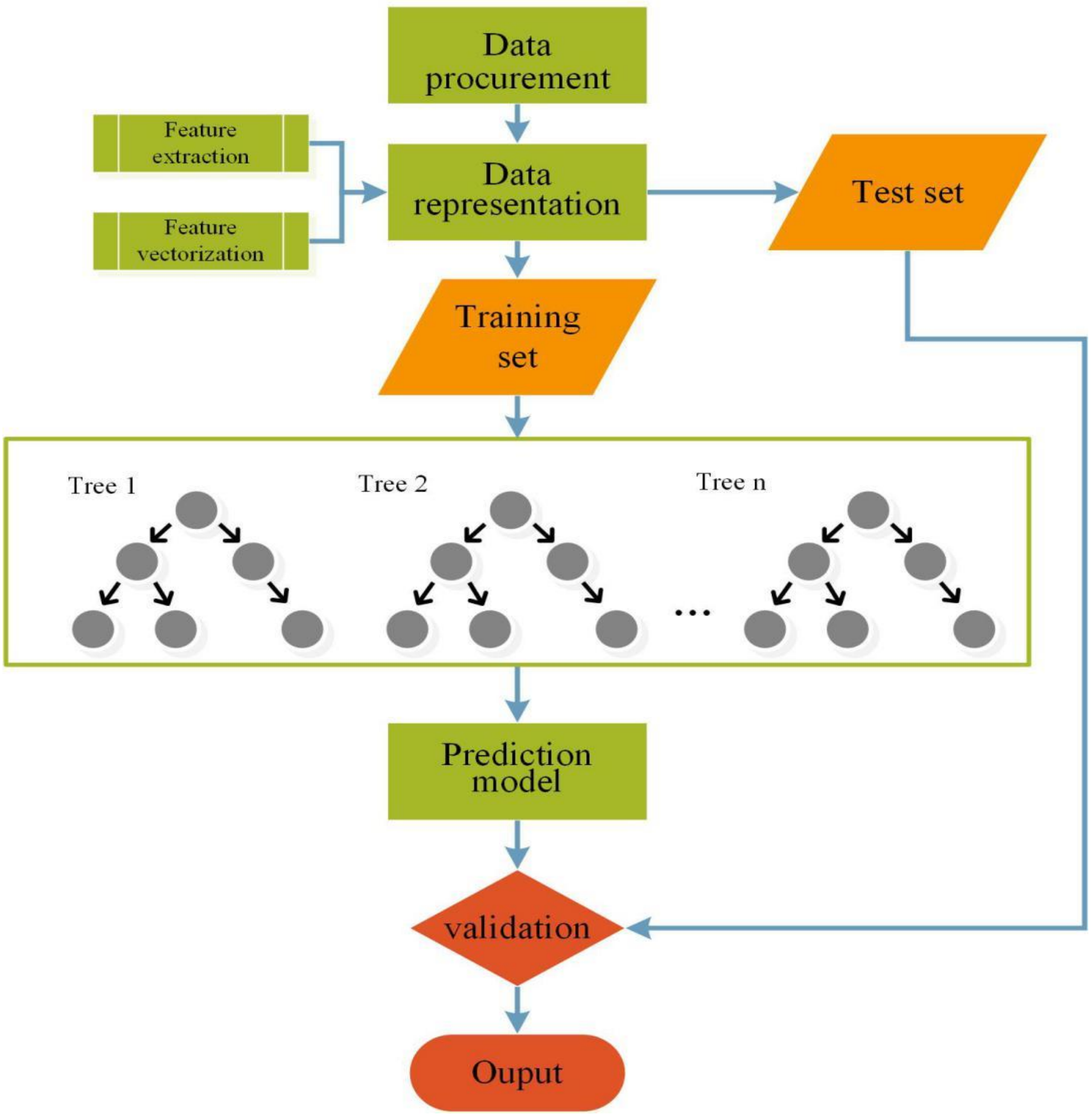

**Fig. 7** Schematics of applying the random forest in studying the importance of different factors.

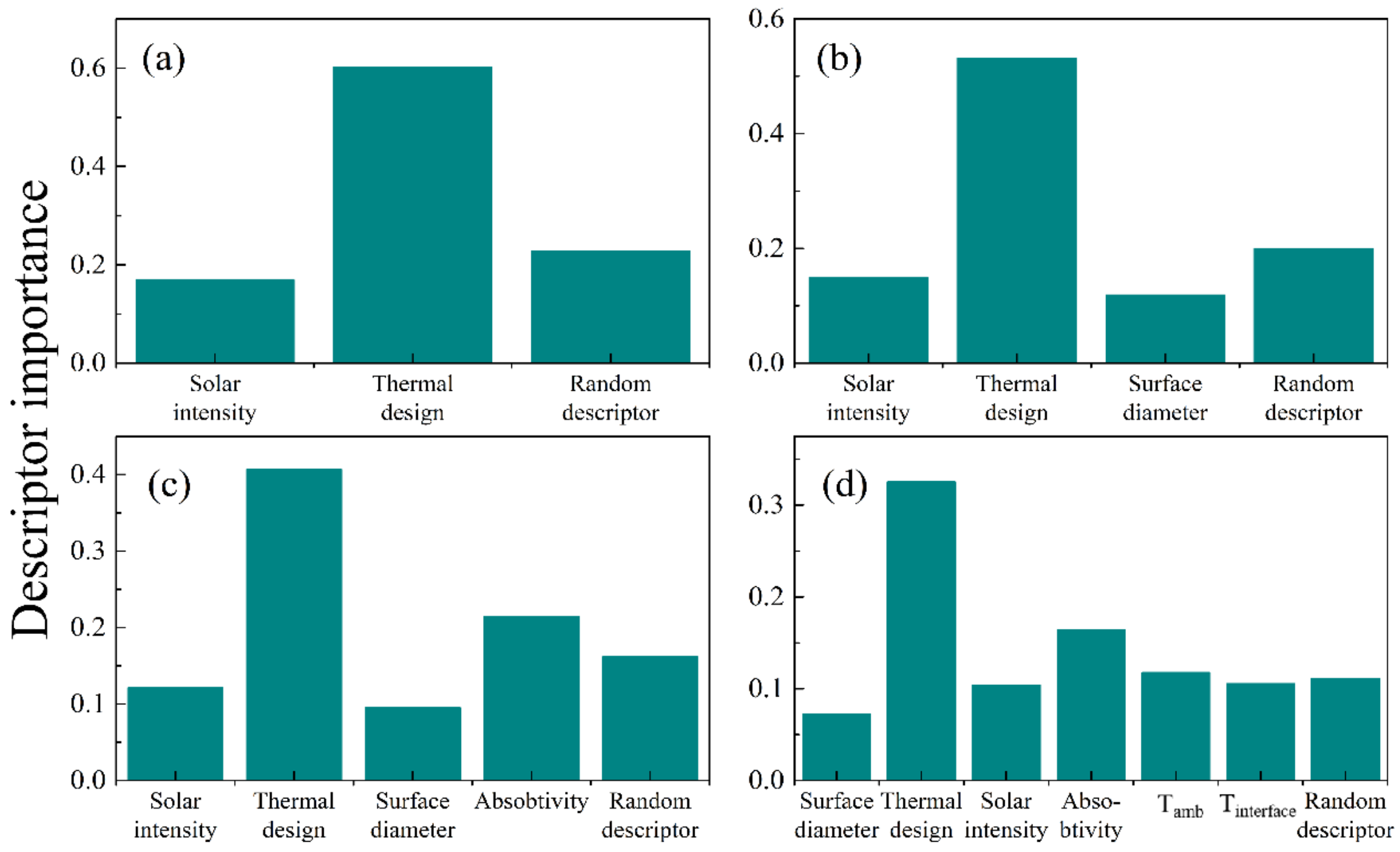

Fig. 8 Results of descriptor importance. The values are normalized as the sum of all values of descriptor importance equals 1. (a)-(d) are the results of using 2, 3,4, and 6 different descriptors, respectively. Surface diameter is the diameter (length) of the evaporation surface. $T_{amb}$ is the temperature of ambient, $T_{interface}$ is the temperature of evaporation interface. Random descriptor is a set of random values and should have no relationship to solar evaporation, which is calculated for comparisons.

## 3.4 Section summary

In summary, interface system exhibits better performance than volumetric system, due to heat localization effect. Porous materials are a better candidate for solar evaporation compared to nanofluids. Various micro/nanostructured porous materials are suitable for solar evaporation, such as paper-based film, artificial aerogel and natural biomaterials. The most efficient materials, especially the hierarchically structured aerogels, exhibit a more than 90% of energy efficiency. Meanwhile, the thermal design shows a significant role in solar evaporation according to calculation results by machine learning. Decrease the heat loss by decreasing the dimension of water channel generally enhances the evaporation efficiency. However, the narrowed water channel may also increase the possibility of salt accumulation, which is quite harmful to seawater based solar evaporation.

# 4. Research gaps between solar still and solar evaporation

## 4.1 Low efficiency of solar still with floatable structure

It is important to prove the effectiveness of applying materials in solar desalination by using solar still with floatable structure. However, there is only a few works carried outdoor experiments by using solar stills. It is found that although the evaporation rate is very high, low productivities and efficiencies are obtained for solar still. The daily productivity of solar still is usually about 1-4 L/day and the corresponding efficiency is below 40% under natural sun, which is far below the evaporation efficiency (usually as high as 60%-90%) [77, 89, 94]. It shows that there is a large research gap between solar evaporation and solar desalination using solar still.

The reasons of lower productivity and efficiency in solar still are usually attributed to the lower solar intensity of natural sunlight, the optical impedance by condensed water drops and low condensation rate of vapor [94a, 95]. In the laboratory, the solar intensity can be maintained at from one sun to as high as more than ten suns. However, the solar intensity of the natural sun, varies from 0 at night to around one sun at noon. The evaporation efficiency will decrease when the solar intensity decreases, due to the lower temperature reached [89]. Meanwhile, when vapor condenses on the transparent cover as a droplet, the reflectance of cover will be increased, which reduces the solar energy received. The optical impedance might be avoided by using a more hydrophilic cover or a vapor-facing-down design as suggested by Liu et al. [95]. On the other hand, the low condensation rate of water vapor is due to the high cover temperature and insufficient condensation area [25a, 96].

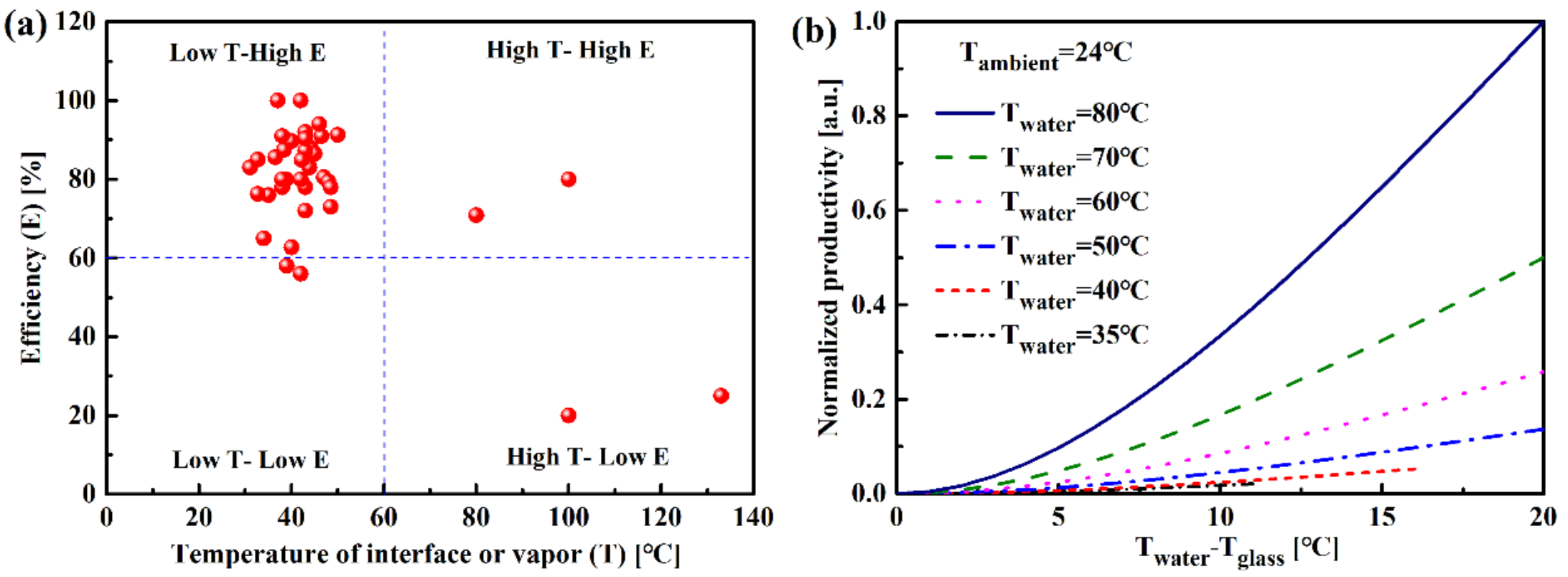

Fig. 9 (a) The efficiency versus the air-water interface temperature (details in Table S2). (b) Calculated productivity at different water temperature and glass temperature.

Besides, there is another essential reason for low productivity and efficiency in solar still, which is usually ignored by most researchers: the low air-water interface temperature. In most of recent works, the temperature of the air-water interface in solar evaporation is usually maintained very low to decrease the heat loss and therefore enhance the energy efficiency. The temperature of the air-water interface under one sun is summarized at Fig. 9 (a) and Table S2. The air-water interface temperature in most works is at around 30-50 °C with the efficiency at around 70-90%. Only few of them reported a temperature of more than 80 °C [64, 68, 97].

But, for solar desalination using solar still, a higher air-water interface temperature will lead to higher productivity and efficiency [14, 98]. A very low air-water interface temperature will result in very low productivity and efficiency in solar still, although a good evaporation rate may be achieved. This is because, the vapor with lower temperature is more difficult to condense on the condensation plate (glass cover) due to the exponentially decreased difference of vapor pressure. Meanwhile, only a few parts of water can be extracted from the vapor when the temperature of vapor is close to that of the condensation plate. To illustrate this point, the productivity of solar still at different water temperature is calculated according to the half empirical equation (1) [7, 99]:

$$Q_e = 0.0144\left[T_w - T_g + \frac{(P_w-P_g)(T_w+273.15)}{268.9\times10^3-P_w}\right](P_w - P_g) \qquad (1)$$

where $Q_e$ is the heat transfer rate between glass and water by condensation, which is proportional to the productivity of solar still. $T_w$ and $T_g$ are the temperature of water evaporation surface and inner surface of glass cover, respectively. $P_w$ is the saturation vapor pressure of water surface at temperature $T_w$. $P_g$ is saturation vapor pressure of the inner surface of glass cover at temperature $T_g$

$$P_w = e^{\left(25.31 - \frac{5144}{T_w + 273.15}\right)} \quad (2)$$

$$P_g = e^{\left(25.31 - \frac{5144}{T_g + 273.15}\right)} \quad (3)$$

For a given temperature difference between the water surface and glass cover, the productivity increase with the increase in water temperature (Fig. 9b). The productivity almost doubled when the water temperature increases 10°C. The productivity of solar still is unable to reach a high value if the water temperature is lower than 40°C, due to the limited heat and mass transfer rate between water and glass cover.

Therefore, materials and thermal designs for both high energy efficiency and high air-water interface temperature are very necessary in the future. It's not only important to increase the productivity of solar still, but also important to some other processes that solar evaporation is involved, such as sterilization [100], latent heat recovery [101] and power generation [54], where high vapor temperature is preferred.

### 4.2 Salt rejecting and durability

Besides the unsatisfied productivity and efficiency of solar still by using foams and films, salt rejecting is another issue that draws a lot of attention. The rapid loss of water in the air-water interface leads to a significant increase in local concentration of salt. When the concentration of salt is saturated, the crystallized salt will reflect the solar irradiation and block the evaporation surface, and then slow down the evaporation[89, 102]. Therefore, whether salt crystallization will occur and affect the evaporation or not must be considered when design materials for solar still.

The reported experiments about salt rejecting can be divided into two categories, cycle test and continuous test, shown in Table 2. In the cycle test, the materials are dried or washed before the next cycle [75, 103], each cycle sustains 0.5-5 hours. Most of the cycle test shows that there is no salt crystallization which might be due to that the salt is removed timely or the good salt rejecting ability of materials. Although salt crystallization is observed in some works, the evaporation is not affected and remains stable [103a, 104]. Nevertheless, it is inconvenient to wash or dry materials circularly in solar still. Therefore, carrying out the long-time continuous test is necessary to further explore the salt rejecting performance of foams and films. It is found that salt crystallization is easy to occur at long-term continues test [102b, 105].

To avoid the salt crystallization, the high concentration of salt on the evaporation interface must be diluted or removed timely. One way is using materials with relatively larger holes (millimeter scale), which ensures that the generated salt particles at the interface can fall to the bulk liquid through holes instead of accumulated on the interface [63a]. The holes also provide dilute solution which exchanges salt ions with the high concentration brine on the interface to avoid salt crystallization, as shown in Fig 10a [106]. Janus structure is another effective strategy to solve salt crystallization (Fig 10b) [102a]. In Janus structure, solar absorption and water pumping, are decoupled into different layers. There is an upper hydrophobic layer (CB/PMMA) for light absorption, and a lower hydrophilic layer (PAN) for pumping water. Therefore, salt can be only deposited in the hydrophilic layer and be dissolved quickly due to continuous water pumping. A more thorough way to overcome salt crystallization on evaporation interface is utilizing contactless structure [97b]. In this method, the solar absorbing structure absorbs solar radiation and re-radiates infrared photons to heat the water (Fig 10 c). The heat transfer between the solar absorbing structure and water is completely by thermal irradiation, instead of heat conduction or convection. Therefore, fouling is entirely avoided due to the physical separation from the water.

However, the evaporation efficiency of the aforementioned three strategy is relatively low (25%-75%). A better way with high efficiency might be enhancing the water absorption ability of materials. For example, due to the free solutions exchange enabled by strong capillary effect, melamine resin sponge achieves an evaporation efficiency of 85% under one sun and 90% under ten suns without salt crystallization after 11 hours of continuing test [105]. Hydrogel can also achieve a water evaporation rate of ~2.6 kg/(m$^2$·h) at ~91% energy efficiency under one sun, without salt crystallization after 100 hours of continuing test [107]. Therefore, improving the water absorption ability of materials might be a very potential way to enhance the salt rejecting performance without losing energy efficiency.

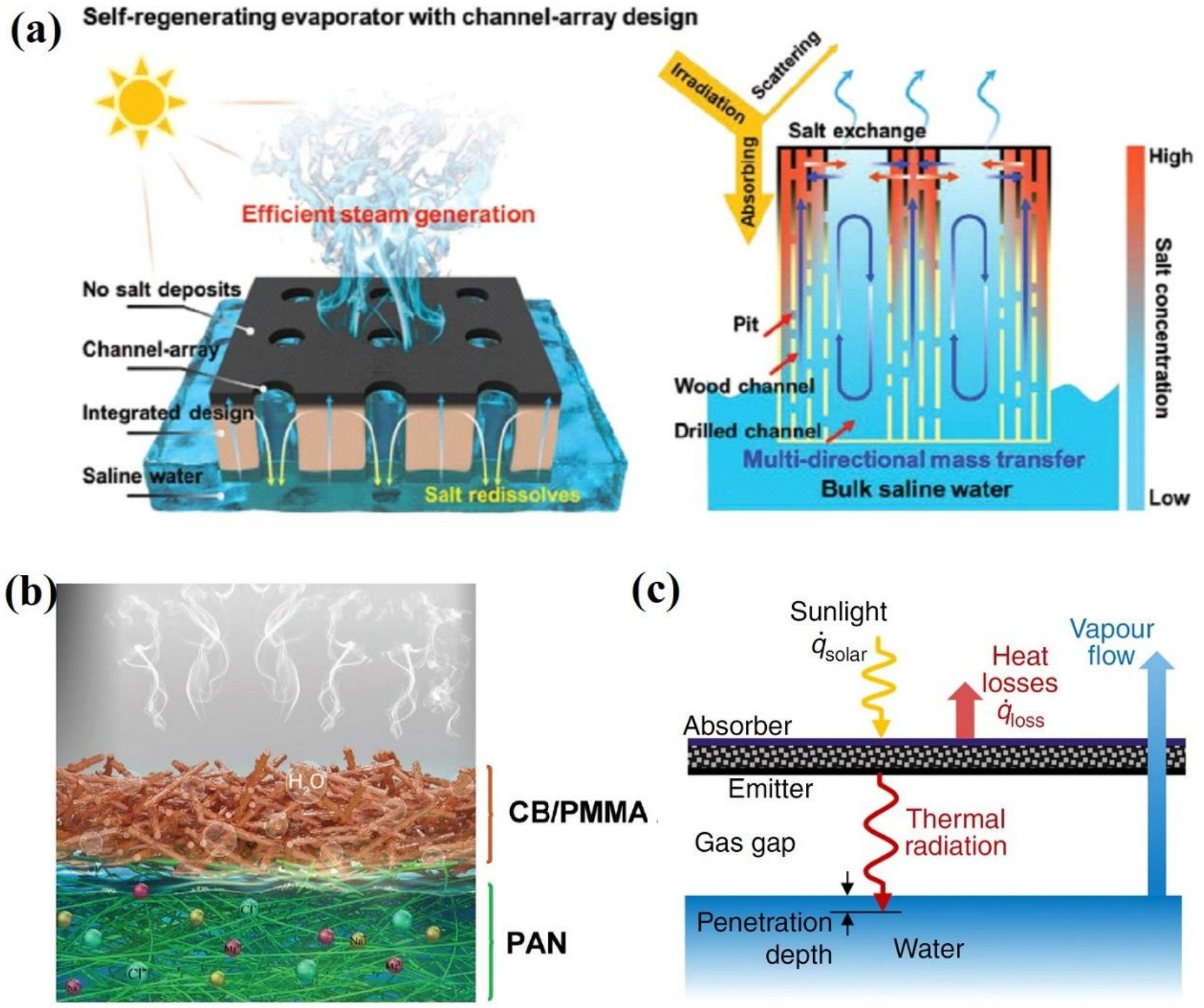

Fig. 10 (a) a diagram showing new design self-regenerating solar evaporator (left), and multipath ways mass transfer in the evaporator (right) [106], Copyright 2019 WILEY-VCH Verlag GmbH & Co. KGaA, Weinheim. (b) Structures of Janus absorber during solar desalination [102a] Copyright 2018 WILEY-VCH Verlag GmbH & Co. KGaA, Weinheim (c) Mechanism of contactless solar evaporation structure. As the absorber heats up, it emits thermal radiation to the water, which is absorbed beneath the water/vapor interface [97b]. Copyright Springer Nature.

Table 2: Durability and salt rejecting performance of materials

| **Categories of test** | **Durations of test** | **Total illuminated time /hours** | **Salt crystallization** | **Water type** | **Solar intensity /kw** |
|---|---|---|---|---|---|
| Circular test | 15 cycles [103b] | Unknown | No | Salt water (3.5 wt.% NaCl) | 1 |
| | 10 cycles [77] | 10 | Yes | Salt water (3.5 wt.% NaCl) | 1 |
| | 20 cycles [75] | Unknown | No | Sea water (salinity 2.75%) | 1 |
| | 10 cycles [55a] | 5 | No | Salt water (3.5 wt.% NaCl) | 1 |
| | 10 cycles [68] | 50 | No | Sea water (East Sea) | 1 |
| | 50 cycles [108] | Unknown | Unknown | Salt water (salinity unknown) | 1 |
| | 15 cycles [104] | 15 | Yes | Salt water (3.5 wt.% NaCl) | 1 |
| | 100 cycles [103a] | 100 | Yes | Salt water (3.5 wt.% NaCl) | 1 |
| | 24 cycles [66a] | 24 | No | Sea water (Bohai Sea) | 2 |
| Continuous test | 20 hours [63a] | 20 | No | Salt water (15 wt.% NaCl) | 1 |
| | 100 hours [106] | 100 | No | Salt water (20 wt.% NaCl) | 1 |
| | 10 hours [92] | 10 | Yes | Sea water (salinity 3.5%) | 1 |
| | 10 hours [74b] | 10 | Unknown | Sea water (East Sea) | 1 |
| | 48 hours [107] | 48 | No | Sea water (salinity 3.5%) | 1 |
| | 10 hours [109] | 10 | No | Salt water (3.5 wt.% NaCl) | 2 |
| | 100 hours [79a] | 100 | Yes | Sea water (Chesapeake Bay) | 5 |
| | 11 hours [105] | 11 | No | Salt water (3.5 wt.% NaCl) | 10 |
| Immersing-Circular test | 28 days [76] | 11 | No | Sea water (Gulf of Mexico) | 1 |
| | 16 days [102a] | 12 | No | Salt water (3.5 wt.% NaCl) | 1 |
| | 30 days [110] | 30 | No | Sea water (Yellow Sea) | 1 |
| | 25 days [111] | Unknown | Unknown | Sea water (East Sea) | 2 |

The durability of materials is another essential point in solar still. The cost and difficulty of maintenance will be increased and the practical application will be limited if materials are required to be replaced frequently, especially for some remote and developing areas. Sea water contains not only salt but also some other contaminants such as minerals, bacteria, and sands, which make the durability of materials more challenging.

Similar to the salt rejecting experiments, most of the durability results are obtained by cycle test or continuous test in salt water or sea water (shown in Table 2). Although the evaporation rate is stable, the duration remains just a couple of days, which is too short to prove the durability of materials. Therefore, several works extended their durability test to several weeks by immersing the samples in seawater [76, 102a, 110-111]. After days or weeks of storage, the samples were taken out and irradiated under one sun irradiation to verify the salt resistance by measuring the evaporation rate. Stable evaporation is observed and no obvious degradation of materials appeared in these tests. Although the samples contact with sea water for weeks, the total time of materials under solar irradiation is only tens of hours. Durability test under both long times of solar irradiation and seawater environment is still lacking.

Therefore, the durability of materials needs more investigations to meet the requirement of practical application in solar still, where solar irradiation and seawater environment might last for months. To simulate the real work condition, it is necessary to carry out long-term experiments by testing the performance of materials in outdoor solar stills or vapor generators for weeks or months.

### 4.3 Effect of solid-liquid interaction on water phase change

Traditional phase change model only focuses on vapor-liquid interaction. The Hertz-Knudsen (HK) equation and its modified forms have proved very effective to predict the phase change rate from droplets [112]. However, when water evaporates from a porous structure, the solid-liquid interaction becomes significant, due to the abundant solid-liquid interfaces and the short distance between the evaporation surface and solid-liquid interface. Many important details which are helpful to manipulate phase change process will be ignored by only focus on vapor-liquid interaction.

Thin-film evaporation theory proves the importance of solid-liquid interaction from the aspect of heat transfer at the microscale level[113]. At the hydrophilic surface of porous foams or films, the water channel is filled with water meniscuses, which is constructed by three regions: (I) adsorbed or non-evaporation region, where water is adsorbed on the graphite due to the high disjoining pressure; (II) thin-film or transition region where effects of long-range molecular forces are felt; (III) intrinsic meniscus region, where the thickness of the water layer increases very fast [114]. In the adsorbed region, water sticks to the graphite tightly, and no mass/heat transfer occurs. Whereas in the thin-film region, the disjoin pressure is weak, while the thickness of the water layer is still thin enough to assure a low thermal resistance. Therefore, the most heat current runs through the thin-film region, hence the fast evaporation and low heat loss [115]. Although the area of thin film region is relatedly small compared to intrinsic meniscus region, the evaporation heat transfer may account for more than 80 % of the total evaporation heat transfer [113a].

At the nano or molecular level, solid-liquid interaction endows fluids unconventional structural properties and dynamical behaviors, which greatly affect the phase change process. For example, the evaporation from water capillary in two-dimensional nanochannels surpasses the traditional Hertz–Knudsen limit by an order of magnitude [116]. The solar evaporation assisted by nanomaterials also shows unconventional phenomenon. The solar evaporation rate under one sun is able to be 110% higher than the traditional limit due to the reduction of phase change enthalpy, which shows a great opportunity for the further improvement of phase change based solar desalination[76, 102b]. One possible reason of the enhanced evaporation is “cluster evaporation”, i.e. water evaporates as molecular cluster, instead of evaporates one molecular by one molecular. However, this explanation is questionable, due to the lack of convincible fundamental theory and experimental observation of water clusters undergoing “evaporation”[95]. Another possible reason is the reduced free energy barrier for evaporation across the liquid-vapor interface [117]. Molecular dynamics reveals that the graphene edges of nanopores can boost the overall evaporative flux by more than 100%. The interaction between the water molecules is weaken by the charged edges of graphene nanopores. This local-field-driven dipole moment destabilizes the H-bond network of water at the interface, and may responsible for the enhancement in the evaporation rate. Nevertheless, many works have to be done to give us a reliable fundamental theory in this field, which is very important for the development of evaporation and solar desalination.

## 5. Conclusions

In this paper, a detailed review of current developments in solar stills with nano/micro materials is presented. Most of recent efforts were interested in improving solar evaporation, which is just one of basic process in solar still. Various materials, such as paper-based film, artificial aerogel and natural biomaterials, were applied to enhance the system efficiency and productivity. By combining heat localization at air-water interface and optimized thermal design, proper materials might achieve an efficiency more than 90% under 1 $kw/m^2$ of solar irradiation. However, the productivity remains low when these materials were applied. This show the large gap between solar still and solar evaporation systems. Except lower natural solar irradiation and optical impedance by condensate droplets, the low water and vapor temperature might be another important reason in these evaporation systems.

Meanwhile, salt rejecting and durability of materials are also two important challenges for the practical application of nano/micro materials. Enhance the water absorption ability of materials might be a very effective way to avoid salt crystallization. Besides the engineering and materials studies, more fundamental research is still missing and should be done to uncover the phase change mechanism at nano and molecular level. Especially, the new understanding on effect of solid-liquid interaction might open a new avenue in solar desalination field.

## Acknowledgments

N.Y. was sponsored by National Natural Science Foundation of China (No. 51576076 and No. 51711540031), Natural Science Foundation of Hubei Province (2017CFA046) and Fundamental Research Funds for the Central Universities (2019kfyRCPY045). M. A. was supported by Natural Science Research Start-up Fund of Shaanxi University of Science and Technology (2018GB-10).

## References

[1] M. Elimelech, W. A. Phillip, *Science* **2011**, 333, 712.

[2] M. S. S. Abujazar, S. Fatihah, E. R. Lotfy, A. E. Kabeel, S. Sharil, *Desalination* **2018**, 425, 94.

[3] L. Sahota, G. N. Tiwari, *Sol Energy* **2016**, 130, 260.

[4] P. Vishwanath Kumar, A. Kumar, O. Prakash, A. K. Kaviti, *Renewable and Sustainable Energy Reviews* **2015**, 51, 153.

[5] M. Gao, L. Zhu, C. K. Peh, G. W. Ho, *Energ Environ Sci* **2019**, 12, 841.

[6] A. E. Kabeel, M. Abdelgaied, M. B. Feddaoui, *Desalination* **2018**, 439, 162.

[7] S. W. Sharshir, N. Yang, G. Peng, A. E. Kabeel, *Appl Therm Eng* **2016**, 100, 267.

[8] K. Kalidasa Murugavel, K. Srithar, *Renew Energ* **2011**, 36, 612.

[9] B. A. K. Abu-Hijleh, H. M. Rababa'h, *Energ Convers Manage* **2003**, 44, 1411.

[10] M. M. Naim, M. A. Abd El Kawi, *Desalination* **2003**, 153, 55.

[11] R. Sathyamurthy, S. A. El-Agouz, P. K. Nagarajan, J. Subramani, T. Arunkumar, D. Mageshbabu, B. Madhu, R. Bharathwaaj, N. Prakash, *Renewable and Sustainable Energy Reviews* **2017**, 77, 1069.

[12] A. Rahmani, A. Boutriaa, *Int J Hydrogen Energ* **2017**, 42, 29047.

[13] Z. M. Omara, A. E. Kabeel, A. S. Abdullah, *Renewable and Sustainable Energy Reviews* **2017**, 68, 638.

[14] P. Durkaieswaran, K. K. Murugavel, *Renewable and Sustainable Energy Reviews* **2015**, 49, 1048.

[15] a) V. Trisaksri, S. Wongwises, *Renewable and Sustainable Energy Reviews* **2007**, 11, 512; b) A. H. Elsheikh, S. W. Sharshir, M. E. Mostafa, F. A. Essa, M. K. Ahmed Ali, *Renewable and Sustainable Energy Reviews* **2018**, 82, 3483.

[16] M. Gao, P. K. N. Connor, G. W. Ho, *Energ Environ Sci* **2016**, 9, 3151.

[17] V.-D. Dao, H.-S. Choi, *Global Challenges* **2018**, 2, 1700094.

[18] X. Yin, Y. Zhang, Q. Guo, X. Cai, J. Xiao, Z. Ding, J. Yang, *ACS Appl Mater Interfaces* **2018**, 10, 10998.

[19] Y. Wang, L. Zhang, P. Wang, *Acs Sustain Chem Eng* **2016**, 4, 1223.

[20] O. Neumann, A. S. Urban, J. Day, S. Lal, P. Nordlander, N. J. Halas, *Acs Nano* **2012**, 7, 42.

[21] H. Ghasemi, G. Ni, A. M. Marconnet, J. Loomis, S. Yerci, N. Miljkovic, G. Chen, *Nat Commun* **2014**, 5, 4449.

[22] G. Ni, N. Miljkovic, H. Ghasemi, X. Huang, S. V. Boriskina, C.-T. Lin, J. Wang, Y. Xu, M. M. Rahman, T. Zhang, G. Chen, *Nano Energy* **2015**, 17, 290.

[23] T. Elango, A. Kannan, K. Kalidasa Murugavel, *Desalination* **2015**, 360, 45.

[24] L. Sahota, G. N. Tiwari, *Desalination* **2016**, 388, 9.

[25] a) S. W. Sharshir, G. Peng, L. Wu, N. Yang, F. A. Essa, A. H. Elsheikh, S. I. T. Mohamed, A. E. Kabeel, *Appl Therm Eng* **2017**, 113, 684; b) S. W. Sharshir, G. Peng, A. H. Elsheikh, E. M. A. Edreis, M. A. Eltawil, T. Abdelhamid, A. E. Kabeel, J. Zang, N. Yang, *Energ Convers Manage* **2018**, 177, 363.

[26] S. W. Sharshir, G. Peng, L. Wu, F. A. Essa, A. E. Kabeel, N. Yang, *Appl Energ* **2017**, 191, 358.

[27] a) A. E. Kabeel, Z. M. Omara, F. A. Essa, *Energ Convers Manage* **2014**, 78, 493; b) A. E. Kabeel, Z. M. Omara, F. A. Essa, *Energ Convers Manage* **2014**, 86, 268.

[28] Z. M. Omara, A. E. Kabeel, F. A. Essa, *Energ Convers Manage* **2015**, 103, 965.
[29] O. Mahian, A. Kianifar, S. Z. Heris, D. Wen, A. Z. Sahin, S. Wongwises, *Nano Energy* **2017**, 36, 134.
[30] L. Sahota, Shyam, G. N. Tiwari, *Energ Convers Manage* **2017**, 135, 308.
[31] G. P. Celata, F. D'Annibale, A. Mariani, S. Sau, E. Serra, R. Bubbico, C. Menale, H. Poth, *Chemical Engineering Research and Design* **2014**, 92, 1616.
[32] R. Bubbico, G. P. Celata, F. D'Annibale, B. Mazzarotta, C. Menale, *Chemical Engineering Research and Design* **2015**, 104, 605.
[33] a) S. M. Shinde, D. M. Kawadekar, P. A. Patil, V. K. Bhojwani, *International Journal of Ambient Energy* **2019**, DOI: 10.1080/01430750.2019.16211981; b) S. Shamshirband, A. Malvandi, A. Karimipour, M. Goodarzi, M. Afrand, D. Petković, M. Dahari, N. Mahmoodian, *Powder Technol* **2015**, 284, 336.
[34] J.-Y. Sha, H.-H. Ge, C. Wan, L.-T. Wang, S.-Y. Xie, X.-J. Meng, Y.-Z. Zhao, *Corros Sci* **2019**, 148, 123.
[35] B. G. N. Muthanna, M. Amara, M. H. Meliani, B. Mettai, Ž. Božić, R. Suleiman, A. A. Sorour, *Eng Fail Anal* **2019**, 102, 293.
[36] N. Sezer, M. A. Atieh, M. Koç, *Powder Technol* **2019**, 344, 404.
[37] R. A. Taylor, P. E. Phelan, R. J. Adrian, A. Gunawan, T. P. Otanicar, *Int J Therm Sci* **2012**, 56, 1.
[38] W. Duangthongsuk, S. Wongwises, *Int J Heat Mass Tran* **2010**, 53, 334.
[39] L. Zhou, S. Zhuang, C. He, Y. Tan, Z. Wang, J. Zhu, *Nano Energy* **2017**, 32, 195.
[40] Y. Wang, M. E. Zaytsev, H. L. The, J. C. Eijkel, H. J. Zandvliet, X. Zhang, D. Lohse, *Acs Nano* **2017**, 11, 2045.
[41] V. Kotaidis, C. Dahmen, G. von Plessen, F. Springer, A. Plech, *J Chem Phys* **2006**, 124, 184702.
[42] D. Lapotko, *Opt Express* **2009**, 17, 2538.
[43] a) Z. Fang, Y. R. Zhen, O. Neumann, A. Polman, F. J. Garcia de Abajo, P. Nordlander, N. J. Halas, *Nano Lett* **2013**, 13, 1736; b) O. Neumann, C. Feronti, A. D. Neumann, A. Dong, K. Schell, B. Lu, E. Kim, M. Quinn, S. Thompson, N. Grady, P. Nordlander, M. Oden, N. J. Halas, *Proc Natl Acad Sci U S A* **2013**, 110, 11677.
[44] N. J. Hogan, A. S. Urban, C. Ayala-Orozco, A. Pimpinelli, P. Nordlander, N. J. Halas, *Nano Lett* **2014**, 14, 4640.
[45] H. Jin, G. Lin, L. Bai, A. Zeiny, D. Wen, *Nano Energy* **2016**, 28, 397.
[46] a) H. Wang, L. Miao, S. Tanemura, *Solar RRL* **2017**, 1, 1600023; b) M. S. Zielinski, J. W. Choi, T. La Grange, M. Modestino, S. M. Hashemi, Y. Pu, S. Birkhold, J. A. Hubbell, D. Psaltis, *Nano Lett* **2016**, 16, 2159.
[47] X. Wang, Y. He, X. Liu, L. Shi, J. Zhu, *Sol Energy* **2017**, 157, 35.
[48] Y.-T. Yen, C.-W. Chen, M. Fang, Y.-Z. Chen, C.-C. Lai, C.-H. Hsu, Y.-C. Wang, H. Lin, C.-H. Shen, J.-M. Shieh, J. C. Ho, Y.-L. Chueh, *Nano Energy* **2015**, 15, 470.
[49] A. Guo, Y. Fu, G. Wang, X. Wang, *Rsc Adv* **2017**, 7, 4815.
[50] A. Zeiny, H. Jin, G. Lin, P. Song, D. Wen, *Renew Energ* **2018**, 122, 443.
[51] X. Wang, G. Ou, N. Wang, H. Wu, *ACS Appl Mater Interfaces* **2016**, 8, 9194.
[52] B. Sharma, M. K. Rabinal, *J Alloy Compd* **2017**, 690, 57.
[53] Y. Fu, T. Mei, G. Wang, A. Guo, G. Dai, S. Wang, J. Wang, J. Li, X. Wang, *Appl Therm Eng*

**2017**, 114, 961.
[54] P. Tao, G. Ni, C. Song, W. Shang, J. Wu, J. Zhu, G. Chen, T. Deng, *Nature Energy* **2018**, DOI: 10.1038/s41560-018-0260-7.
[55] a) Z. Wang, Q. Ye, X. Liang, J. Xu, C. Chang, C. Song, W. Shang, J. Wu, P. Tao, T. Deng, *J Mater Chem A* **2017**, 5, 16359; b) X. Liu, B. Hou, G. Wang, Z. Cui, X. Zhu, X. Wang, *J Mater Res* **2018**, 33, 674.
[56] J. Lou, Y. Liu, Z. Wang, D. Zhao, C. Song, J. Wu, N. Dasgupta, W. Zhang, D. Zhang, P. Tao, W. Shang, T. Deng, *ACS Appl Mater Interfaces* **2016**, 8, 14628.
[57] G. Wang, Y. Fu, X. Ma, W. Pi, D. Liu, X. Wang, *Carbon* **2017**, 114, 117.
[58] Y. Liu, S. Yu, R. Feng, A. Bernard, Y. Liu, Y. Zhang, H. Duan, W. Shang, P. Tao, C. Song, T. Deng, *Adv Mater* **2015**, 27, 2768.
[59] a) Z. Wang, Y. Liu, P. Tao, Q. Shen, N. Yi, F. Zhang, Q. Liu, C. Song, D. Zhang, W. Shang, T. Deng, *Small* **2014**, 10, 3234; b) Y. Liu, J. Lou, M. Ni, C. Song, J. Wu, N. P. Dasgupta, P. Tao, W. Shang, T. Deng, *ACS Appl Mater Interfaces* **2016**, 8, 772.
[60] K. Bae, G. Kang, S. K. Cho, W. Park, K. Kim, W. J. Padilla, *Nat Commun* **2015**, 6, 10103.
[61] J. Wang, Z. Liu, X. Dong, C.-E. Hsiung, Y. Zhu, L. Liu, Y. Han, *J Mater Chem A* **2017**, 5, 6860.
[62] F. Chen, A. S. Gong, M. Zhu, G. Chen, S. D. Lacey, F. Jiang, Y. Li, Y. Wang, J. Dai, Y. Yao, J. Song, B. Liu, K. Fu, S. Das, L. Hu, *Acs Nano* **2017**, 11, 4275.
[63] a) V. Kashyap, A. Al-Bayati, S. M. Sajadi, P. Irajizad, S. H. Wang, H. Ghasemi, *J Mater Chem A* **2017**, 5, 15227; b) X. Wang, Y. He, X. Liu, G. Cheng, J. Zhu, *Appl Energ* **2017**, 195, 414; c) M. Chen, Y. Wu, W. Song, Y. Mo, X. Lin, Q. He, B. Guo, *Nanoscale* **2018**, 10, 6186; d) M. Kaur, S. Ishii, S. L. Shinde, T. Nagao, *Acs Sustain Chem Eng* **2017**, 5, 8523.
[64] Y. Ito, Y. Tanabe, J. Han, T. Fujita, K. Tanigaki, M. Chen, *Adv Mater* **2015**, 27, 4302.
[65] R. Li, L. Zhang, L. Shi, P. Wang, *Acs Nano* **2017**, 11, 3752.
[66] a) L. Zhou, Y. Tan, J. Wang, W. Xu, Y. Yuan, W. Cai, S. Zhu, J. Zhu, *Nat Photonics* **2016**, 10, 393; b) L. Zhou, Y. Tan, D. Ji, B. Zhu, P. Zhang, J. Xu, Q. Gan, Z. Yu, J. Zhu, *Sci. Adv.* **2016**, 2, e1501227
[67] X. Yang, Y. Yang, L. Fu, M. Zou, Z. Li, A. Cao, Q. Yuan, *Adv Funct Mater* **2018**, 28, 1704505.
[68] G. Zhu, J. Xu, W. Zhao, F. Huang, *ACS Appl Mater Interfaces* **2016**, 8, 31716.
[69] P. Fan, H. Wu, M. Zhong, H. Zhang, B. Bai, G. Jin, *Nanoscale* **2016**, 8, 14617.
[70] P. Zhang, J. Li, L. Lv, Y. Zhao, L. Qu, *Acs Nano* **2017**, 11, 5087.
[71] L. Tian, J. Luan, K. K. Liu, Q. Jiang, S. Tadepalli, M. K. Gupta, R. R. Naik, S. Singamaneni, *Nano Lett* **2016**, 16, 609.
[72] a) F. Jiang, H. Liu, Y. Li, Y. Kuang, X. Xu, C. Chen, H. Huang, C. Jia, X. Zhao, E. Hitz, Y. Zhou, R. Yang, L. Cui, L. Hu, *ACS Appl Mater Interfaces* **2018**, 10, 1104; b) Q. Jiang, H. Gholami Derami, D. Ghim, S. Cao, Y.-S. Jun, S. Singamaneni, *J Mater Chem A* **2017**, 5, 18397.
[73] Y. Yang, R. Zhao, T. Zhang, K. Zhao, P. Xiao, Y. Ma, P. M. Ajayan, G. Shi, Y. Chen, *Acs Nano* **2018**, 12, 829.
[74] a) X. Hu, W. Xu, L. Zhou, Y. Tan, Y. Wang, S. Zhu, J. Zhu, *Adv Mater* **2017**, 29; b) P. Zhang, Q. Liao, T. Zhang, H. Cheng, Y. Huang, C. Yang, C. Li, L. Jiang, L. Qu, *Nano Energy* **2018**, 46, 415.
[75] H. Ren, M. Tang, B. Guan, K. Wang, J. Yang, F. Wang, M. Wang, J. Shan, Z. Chen, D. Wei, H.

Peng, Z. Liu, *Adv Mater* **2017**, 29, 1702590.
[76] F. Zhao, X. Zhou, Y. Shi, X. Qian, M. Alexander, X. Zhao, S. Mendez, R. Yang, L. Qu, G. Yu, *Nat Nanotechnol* **2018**, 13, 489.
[77] Z. Liu, H. Song, D. Ji, C. Li, A. Cheney, Y. Liu, N. Zhang, X. Zeng, B. Chen, J. Gao, Y. Li, X. Liu, D. Aga, S. Jiang, Z. Yu, Q. Gan, *Global Chall* **2017**, 1, 1600003.
[78] X. Gao, H. Ren, J. Zhou, R. Du, C. Yin, R. Liu, H. Peng, L. Tong, Z. Liu, J. Zhang, *Chem Mater* **2017**, 29, 5777.
[79] a) M. Zhu, Y. Li, G. Chen, F. Jiang, Z. Yang, X. Luo, Y. Wang, S. D. Lacey, J. Dai, C. Wang, C. Jia, J. Wan, Y. Yao, A. Gong, B. Yang, Z. Yu, S. Das, L. Hu, *Adv Mater* **2017**, 29; b) G. Xue, K. Liu, Q. Chen, P. Yang, J. Li, T. Ding, J. Duan, B. Qi, J. Zhou, *ACS Appl Mater Interfaces* **2017**, 9, 15052; c) N. Xu, X. Hu, W. Xu, X. Li, L. Zhou, S. Zhu, J. Zhu, *Adv Mater* **2017**, 29.
[80] H. Liu, C. Chen, G. Chen, Y. Kuang, X. Zhao, J. Song, C. Jia, X. Xu, E. Hitz, H. Xie, S. Wang, F. Jiang, T. Li, Y. Li, A. Gong, R. Yang, S. Das, L. Hu, *Adv Energy Mater* **2018**, 8, 1701616.
[81] K. K. Liu, Q. Jiang, S. Tadepalli, R. Raliya, P. Biswas, R. R. Naik, S. Singamaneni, *ACS Appl Mater Interfaces* **2017**, 9, 7675.
[82] C. Chen, Y. Li, J. Song, Z. Yang, Y. Kuang, E. Hitz, C. Jia, A. Gong, F. Jiang, J. Y. Zhu, B. Yang, J. Xie, L. Hu, *Adv Mater* **2017**, 29.
[83] M. Zhu, Y. Li, F. Chen, X. Zhu, J. Dai, Y. Li, Z. Yang, X. Yan, J. Song, Y. Wang, E. Hitz, W. Luo, M. Lu, B. Yang, L. Hu, *Adv Energy Mater* **2018**, 8, 1701028.
[84] X. Wu, G. Y. Chen, W. Zhang, X. Liu, H. Xu, *Advanced Sustainable Systems* **2017**, 1, 1700046.
[85] C. Jia, Y. Li, Z. Yang, G. Chen, Y. Yao, F. Jiang, Y. Kuang, G. Pastel, H. Xie, B. Yang, S. Das, L. Hu, *Joule* **2017**, 1, 588.
[86] J. Liu, Q. Liu, D. Ma, Y. Yuan, J. Yao, W. Zhang, H. Su, Y. Su, J. Gu, D. Zhang, *J Mater Chem A* **2019**, 7, 9034.
[87] S. Zhuang, L. Zhou, W. Xu, N. Xu, X. Hu, X. Li, G. Lv, Q. Zheng, S. Zhu, Z. Wang, J. Zhu, *Adv Sci (Weinh)* **2018**, 5, 1700497.
[88] X. Li, W. Xu, M. Tang, L. Zhou, B. Zhu, S. Zhu, J. Zhu, *Proc Natl Acad Sci U S A* **2016**, 113, 13953.
[89] G. Ni, S. H. Zandavi, S. M. Javid, S. V. Boriskina, T. A. Cooper, G. Chen, *Energ Environ Sci* **2018**, 11, 1510.
[90] Y. Li, T. Gao, Z. Yang, C. Chen, Y. Kuang, J. Song, C. Jia, E. M. Hitz, B. Yang, L. Hu, *Nano Energy* **2017**, 41, 201.
[91] X. Li, R. Lin, G. Ni, N. Xu, X. Hu, B. Zhu, G. Lv, J. Li, S. Zhu, J. Zhu, *Natl Sci Rev* **2018**, 5, 70.
[92] P.-F. Liu, L. Miao, Z. Deng, J. Zhou, H. Su, L. Sun, S. Tanemura, W. Cao, F. Jiang, L.-D. Zhao, *Materials Today Energy* **2018**, 8, 166.
[93] a) L. Breiman, *Mach Learn* **2001**, 45, 5; b) K. T. Butler, D. W. Davies, H. Cartwright, O. Isayev, A. Walsh, *Nature* **2018**, 559, 547; c) R. Ma, D. Huang, T. Zhang, T. Luo, *Chem Phys Lett* **2018**, 704, 49; d) C. Maltecca, D. Lu, C. Schillebeeckx, N. P. McNulty, C. Schwab, C. Shull, F. Tiezzi, *Scientific Reports* **2019**, 9, 6574; e) S. Subramanian, S. Huq, T. Yatsunenko, R. Haque, M. Mahfuz, M. A. Alam, A. Benezra, J. DeStefano, M. F. Meier, B. D. Muegge, M. J. Barratt, L. G. VanArendonk, Q. Zhang, M. A. Province, W. A. Petri Jr, T. Ahmed, J. I. Gordon, *Nature* **2014**, 510, 417.
[94] a) L. Yi, S. Ci, S. Luo, P. Shao, Y. Hou, Z. Wen, *Nano Energy* **2017**, 41, 600; b) L. Zhang, B.

Tang, J. Wu, R. Li, P. Wang, *Adv Mater* **2015**, 27, 4889.
[95] H. Liu, Z. Huang, K. Liu, X. Hu, J. Zhou, *Adv Energy Mater* **2019**, DOI: 10.1002/aenm.2019003101900310.
[96] R. Bhardwaj, M. V. ten Kortenaar, R. F. Mudde, *Appl Energ* **2015**, 154, 480.
[97] a) G. Ni, G. Li, Svetlana V. Boriskina, H. Li, W. Yang, T. Zhang, G. Chen, *Nature Energy* **2016**, 1, 16126; b) T. A. Cooper, S. H. Zandavi, G. W. Ni, Y. Tsurimaki, Y. Huang, S. V. Boriskina, G. Chen, *Nat Commun* **2018**, 9, 5086.
[98] a) A. E. Kabeel, Z. M. Omara, M. M. Younes, *Renewable and Sustainable Energy Reviews* **2015**, 46, 178; b) A. K. Kaviti, A. Yadav, A. Shukla, *Renewable and Sustainable Energy Reviews* **2016**, 54, 429; c) K. H. Nayi, K. V. Modi, *Renewable and Sustainable Energy Reviews* **2018**, 81, 136; d) S. W. Sharshir, A. H. Elsheikh, G. Peng, N. Yang, M. O. A. El-Samadony, A. E. Kabeel, *Renewable and Sustainable Energy Reviews* **2017**, 73, 521.
[99] V. Manikandan, K. Shanmugasundaram, S. Shanmugan, B. Janarthanan, J. Chandrasekaran, *Renewable and Sustainable Energy Reviews* **2013**, 20, 322.
[100] a) Y. Zhang, D. Zhao, F. Yu, C. Yang, J. Lou, Y. Liu, Y. Chen, Z. Wang, P. Tao, W. Shang, J. Wu, C. Song, T. Deng, *Nanoscale* **2017**, 9, 19384; b) J. Li, M. Du, G. Lv, L. Zhou, X. Li, L. Bertoluzzi, C. Liu, S. Zhu, J. Zhu, *Adv Mater* **2018**, DOI: 10.1002/adma.201805159e1805159.
[101] G. Xue, Q. Chen, S. Lin, J. Duan, P. Yang, K. Liu, J. Li, J. Zhou, *Global Challenges* **2018**, 2, 1800001.
[102] a) W. Xu, X. Hu, S. Zhuang, Y. Wang, X. Li, L. Zhou, S. Zhu, J. Zhu, *Adv Energy Mater* **2018**, 8, 1702884; b) F. Liu, B. Zhao, W. Wu, H. Yang, Y. Ning, Y. Lai, R. Bradley, *Adv Funct Mater* **2018**, 28, 1803266.
[103] a) N. Han, K. Liu, X. Zhang, M. Wang, P. Du, Z. Huang, D. Zhou, Q. Zhang, T. Gao, Y. Jia, L. Luo, J. Wang, X. Sun, *Sci Bull* **2019**, 64, 391; b) F. Gong, H. Li, W. Wang, J. Huang, D. Xia, J. Liao, M. Wu, D. V. Papavassiliou, *Nano Energy* **2019**, 58, 322.
[104] H. Su, J. Zhou, L. Miao, J. Shi, Y. Gu, P. Wang, Y. Tian, X. Mu, A. Wei, L. Huang, S. Chen, Z. Deng, *Sustainable Materials and Technologies* **2019**, 20, e00095.
[105] Y. Liu, Z. Liu, Q. Huang, X. Liang, X. Zhou, H. Fu, Q. Wu, J. Zhang, W. Xie, *J Mater Chem A* **2019**, 7, 2581.
[106] Y. Kuang, C. Chen, S. He, E. M. Hitz, Y. Wang, W. Gan, R. Mi, L. Hu, *Adv Mater* **2019**, DOI: 10.1002/adma.201900498e1900498.
[107] Y. Guo, F. Zhao, X. Zhou, Z. Chen, G. Yu, *Nano Lett* **2019**, 19, 2530.
[108] P. Qiao, J. Wu, H. Li, Y. Xu, L. Ren, K. Lin, W. Zhou, *ACS Appl Mater Interfaces* **2019**, 11, 7066.
[109] P.-F. Liu, L. Miao, Z. Deng, J. Zhou, Y. Gu, S. Chen, H. Cai, L. Sun, S. Tanemura, *Appl Energ* **2019**, 241, 652.
[110] C. Li, D. Jiang, B. Huo, M. Ding, C. Huang, D. Jia, H. Li, C.-Y. Liu, J. Liu, *Nano Energy* **2019**, 60, 841.
[111] Q. Gan, T. Zhang, R. Chen, X. Wang, M. Ye, *Acs Sustain Chem Eng* **2019**, 7, 3925.
[112] A. H. Persad, C. A. Ward, *Chem Rev* **2016**, 116, 7727.
[113] a) R. Ranjan, J. Y. Murthy, S. V. Garimella, *Int J Heat Mass Tran* **2011**, 54, 169; b) F. Su, H. Ma, X. Han, H. H. Chen, B. Tian, *Appl Phys Lett* **2012**, 101, 113702; c) J. Yang, Y. Pang, W. Huang, S. K. Shaw, J. Schiffbauer, M. A. Pillers, X. Mu, S. Luo, T. Zhang, Y. Huang, G. Li, S.

Ptasinska, M. Lieberman, T. Luo, *Acs Nano* **2017**, 11, 5510.
[114] H. Wang, S. V. Garimella, J. Y. Murthy, *Int J Heat Mass Tran* **2007**, 50, 3933.
[115] G. Peng, H. Ding, S. W. Sharshir, X. Li, H. Liu, D. Ma, L. Wu, J. Zang, H. Liu, W. Yu, H. Xie, N. Yang, *Appl Therm Eng* **2018**, 143, 1079.
[116] Y. Li, M. A. Alibakhshi, Y. Zhao, C. Duan, *Nano Lett* **2017**, 17, 4813.
[117] S. Feng, Z. Xu, *Nanotechnology* **2019**, 30, 165401.

# Supporting information

## Micro/nanomaterials for improving solar still and solar evaporation - A review

Guilong Peng[1,#], Swellam W. Sharshir[1,2,3, #], Yunpeng Wang[1], Meng An[4], A.E. Kabeel[5], Jianfeng Zang[2], Lifa Zhang[6], Nuo Yang[1*]

1 State Key Laboratory of Coal Combustion, School of Energy and Power Engineering, Huazhong University of Science and Technology, Wuhan 430074, China

2 School of Optical and Electronic Information, Huazhong University of Science and Technology, Wuhan 430074, China

3 Mechanical Engineering Department, Faculty of Engineering, Kafrelsheikh University, Kafrelsheikh, Egypt

4 College of Mechanical and Electrical Engineering, Shaanxi University of Science and Technology, Xi'an, 710021, China

5 Mechanical Power Engineering Department, Faculty of Engineering, Tanta University, Tanta, Egypt

6 Center for Quantum Transport and Thermal Energy Science, School of Physics and Technology, Nanjing Normal University, Nanjing, 210023, China

[#]Guilong Peng and Swellam W. Sharshir contribute equally on this work.

*Corresponding email: nuo@hust.edu.cn

## Method

Random forest (RF) is a typical ensemble method, which combines multiple decision trees (DT) into one model to improve performance [1]. Given an initial dataset S, a random forest model consist of K decision trees can be established as follows: First, the bootstrap resampling method [2] is used to randomly generate K sets of data from the initial dataset S. Second, K decision trees will be grown based on K sets by specific random selection algorithm in each node [3]. Finally, the finally prediction is made by a weighted vote of all decision trees.

In the present study, classification and regression trees (CART) [4] were used to construct the forest. For each tree in the forest, the Gini index defined in Eq (1) is used to evaluate the purity of a node in the tree. If all samples of a node belong to the same class, the Gini index equals to 0.

$$G(t) = 1 - \sum_{i \in I} p_i^2 \qquad (1)$$

Where G(t) is the Gini index of node t, $p_i$ is the relative frequency of class i in the node t.

After model construction, RF can calculate importance of certain descriptor. For a single tree, a descriptor's importance is defined as the sum of Gini index reduction over all nodes in which the certain descriptor is chose to split [5]. The final descriptor importance is averaged among all trees.

## Dataset

86 Experimental data used in current study were acquired by collecting from articles since 2014. Due to the lack of details in original articles, there are some missing data of Surface diameter, Absorptivity, $T_{amb}$ and $T_{interface}$. The amount of missing data is 16, 22, 33, 24, respectively. The method of Mean Completer was used for filling missing data. Each descriptor was divided into three labels. The details of data representation are listed in Table S1.

**Table S1** Details of data representation

| Descriptors | Classification | Labels | Number of samples |
|---|---|---|---|
| Solar intensity | 1 Kw | 0 | 51 |
| | 1-10 Kw | 1 | 25 |
| | >10 Kw | 2 | 10 |
| Thermal design | 3D interface | 0 | 53 |
| | 2D\1D interface | 1 | 21 |
| | Volumetric | 2 | 12 |
| Surface diameter | <3 cm | 0 | 30 |
| | 3-4 cm | 1 | 29 |
| | >4 cm | 2 | 29 |
| Absorptivity | <0.95 | 0 | 23 |
| | 0.95 | 1 | 31 |
| | >0.95 | 2 | 34 |
| $T_{amb}$ | <24℃ | 0 | 21 |
| | 24-25℃ | 1 | 40 |
| | >25℃ | 2 | 27 |
| $T_{interface}$ | <50℃ | 0 | 37 |
| | 50-70℃ | 1 | 31 |
| | >70℃ | 2 | 20 |
| Efficiency | <75 % | 0 | 28 |
| | 75 %~85 % | 1 | 31 |
| | >85% | 2 | 28 |

**Table S2** Dataset of machine learning

| Ref | Categories | I_solar (kW/m²) | $T_{amb}$/RH (℃/ ) | Water type | Thermal design | Absorptivity (%) | $T_{interface/vapor}$ (℃) | E_water (kg/(m²·h) | Ev_rate (kg/(m²·h) | Ev_dark (kg/(m²·h) | Calibrated with Ev_dark | E.F. (%) |
|---|---|---|---|---|---|---|---|---|---|---|---|---|
| [6] | floating particles | 1.355 | 25 /50 | salt water (3.5%) | 3D interface | — | 68 | 1 | 2.3 | — | Unknown | — |
| [7] | floating particles | 1 | 25 /50 | salt water (3.5%) | 3D interface | — | 46 | 0.56 | 1.28 | — | Unknown | — |
| [8] | nanofluid | 10 | 24 /— | water | volumetric | 99.5 (350-1900nm) | 95 | — | 10.9 | — | Unknown | 69 |
| [9] | nanofluid | 353 | — | water | volumetric | — | 76 | — | — | — | — | 35.6 |
| [10] | nanofluid | 0.8 | 21.4 /41 | water | volumetric | — | — | — | — | — | — | 51 |
| [11] | nanofluid | 0.8 | 20 /40 | water | volumetric | — | 33.4 | — | — | — | — | — |
| [12] | nanofluid | 206 | 35 /— | water | volumetric | — | 100 | — | — | — | — | 64.1 |
| [13] | nanofluid | 10 | — | water | volumetric | 90 (200-2500nm) | — | — | 8.5 | — | Unknown | 46.8 |
| [14] | nanofluid | 1 | 26 /— | salt water (3.5%) | volumetric | 95 (220-2000nm) | — | — | 1.12 | — | Yes | 70 |
| [15] | nanofluid | 11 | 20.6 /20 | water | volumetric | — | — | — | — | — | — | 69 |
| [16] | nanofluid | 280 | — | water | volumetric | — | >100 | — | — | — | — | 84 |
| [17] | nanofluid | 16.7 | 25 /55 | water | volumetric | — | — | 4.4 | 15.74 | — | Unknown | 59.6 |
| [18] | nanofluid | 1 | 28 /49 | water | volumetric | — | — | — | 0.86 | — | Unknown | 53.6 |
| [19] | nanofluid | 10 | 25 /25 | water | volumetric | — | — | — | — | — | — | 60.3 |
| [20] | nanofluid | 1 | — | water | volumetric | — | — | — | 1.1 | — | Unknown | 66 |
| [21] | film | 50.9 | 30.4 /— | water | 3D interface | — | 57.8 | — | — | 0.114 | — | 44 |
| [22] | film | 20 | | water | 3D interface | 91 (400-2500nm) | 80 | — | 15.95 | 1.13 | Yes | 57 |
| [23] | film | 1 | 24 /14 | water | 3D interface | 97 (250-2000nm) | 100 | 0.36 | 1.5 | — | Unknown | 80 |
| [24] | film | 4.5 | 27 /— | water | 3D interface | 87 (400-800nm) | 80 | 0.78 | 4.9 | — | Unknown | 77.8 |
| [25] | film | 1 | 22 /50 | water | 3D interface | — | 39 | 0.39 | 0.92 | 0.088 | Yes | 58 |
| [26] | film | 1 | — | water | Bottom Heating | 95 (200-2000nm) | — | — | — | — | — | 60 |
| [27] | film | 1 | 22 /— | salt water (3.5%) | 3D interface | — | — | 0.59 | 1.01 | — | Unknown | 63.6 |
| [28] | film | 6.7 | — | water | 3D interface | — | 62 | — | — | — | — | — |
| [29] | film | 1 | — | water | 3D interface | 98 (200-800nm) | — | 0.4 | 1.31 | 0.08 | Unknown | 82 |
| [30] | film | 1 | — | water | 3D interface | — | — | — | — | 0.117 | — | 77.1 |
| [31] | film | 4 | 24 /42 | water | 3D interface | 99 (200nm-10μm) | 65 | 1.55 | 5.6 | — | Yes | 90 |
| [32] | film | 4 | 24 /48 | water | 3D interface | 96 (400-1500nm) | 68 | 2.37 | 5.7 | — | Yes | 88 |
| [33] | film | 1 | — | water | 3D interface | 90 (400-1800nm) | 80 | 0.59 | 1.13 | — | Unknown | 70.9 |
| [34] | film | 10 | — | water | 3D interface | 98 (300-1200nm) | 108 | 2.48 | 11.22 | — | Unknown | 81 |
| [35] | film | 5 | — | water | 3D interface | 85 (200-2000nm) | 61 | 3 | 4.95 | — | Unknown | 62 |
| [36] | film | 1 | 26 /20 | water | 3D interface | 94 (300-1500nm) | — | 0.42 | 1.55 | 0.1 | Yes | 91 |
| [37] | film | 1 | — | water | 3D interface | — | 40 | 0.58 | 1.01 | — | Unknown | 62.7 |
| [38] | film | 1 | — | water | 3D interface | >97 (350-1450nm) | 35 | 0.44 | 1.18 | — | Unknown | 76 |
| [39] | film | 4 | 25 /— | water | 3D interface | 95 (300-1200nm) | 45.4 | 0.89 | 4 | — | Unknown | 71.8 |
| [40] | film | 1.95 | — | water | 3D interface | — | 50 | 0.45 | 2.14 | — | Unknown | 72.5 |
| [41] | film | 10 | — | water | 3D interface | 90 (400-1500nm) | — | 2.8 | 11.8 | — | Unknown | 85 |
| [42] | film | 1 | 22 /60 | water | 3D interface | 94 (400-2500nm) | — | 0.1 | 0.47 | 0.06 | Yes | 48 |
| [43] | film | 1 | 23 /55 | water | 3D interface | 90.2 (250-2500nm) | — | 0.509 | 1.24 | — | Unknown | 77.5 |
| [44] | film | 15 | 22 /36 | water | 3D interface | 99 (280-820nm) | 100 | 2.1 | 21 | 0.16 | Yes | 90 |
| [45] | film | 4 | 25 /22 | water | 1D interface | 98 (400-2500) | 73.5 | 2.4 | 6.01 | 0.24 | Yes | 94 |
| [46] | film | 2.94 | 25 /50 | water | 3D interface | — | 98.1 | 1.83 | 3.81 | — | Unknown | 81.4 |
| [47] | film | 1 | — | water | 3D interface | 67.4 (400-1000nm) | 31 | 0.1 | 1.42 | — | Yes | 83 |
| [48] | film | 1 | 25 /— | water | 3D interface | 90-96 (400-2500) | 38.5 | 0.51 | 1.43 | 0.24 | Yes | 90.4 |
| [49] | film | 6 | — | water | 3D interface | 97 (200-2500nm) | 60 | 3.96 | 8.24 | — | Unknown | 83 |
| [50] | film | 5 | — | water | 3D interface | 95 (300-2500nm) | 50 | 1.3 | 6.6 | — | Unknown | 91.5 |
| [51] | foam | 10 | 24 /31 | water | 3D interface | 97 (250-2250nm) | 100 | 5 | 12 | 0.074 | Yes | 85 |
| [52] | foam | 1.7 | — | water | 3D interface | 94 (300-800nm) | 48 | 0.6 | 1.53 | — | Unknown | 60 |
| [53] | foam | 10 | — | water | 3D interface | 96 (400-1100nm) | 100 | 4.8 | 11.8 | — | Unknown | 83 |
| [54] | foam | 1 | — | water | 2D Interface | 94(250-2500nm) | 38.8 | 0.24 | 1.45 | 0.065 | Yes | 80 |
| [55] | foam | 1 | 20 /— | water | 3D interface | 93 (200-2000nm) | 100 | — | — | — | — | 20 |
| [56] | foam | 24 | — | water | 3D interface | 97 (250-2500nm) | 100 | — | — | — | — | — |
| [57] | foam | 51 | — | water | 3D interface | — | 100 | — | 51.8 | — | Unknown | 76.3 |
| [58] | foam | 10 | 26 /52.5 | water | 3D interface | 95 (250-2500nm) | — | — | 14.36 | — | Unknown | 82.7 |
| [59] | foam | 1 | 25 /45 | water | 3D interface | 92 (200-2500nm) | — | 0.5 | 1.622 | — | Unknown | 83 |
| [60] | foam | 10 | 20 /— | water | 3D interface | 92 (250-2500nm) | 92 | 2.9 | 12.1 | 0.12 | Yes | 86.7 |
| [61] | foam | 1 | 25 /— | water | 3D interface | 98(450nm-750nm) | 43 | 0.46 | 1.13 | — | Unknown | 78 |
| [62] | foam | 1 | 20.5 /47 | water | 3D interface | — | 42 | — | — | — | — | 80 |
| [63] | foam | 1 | — | water | 2D interface | — | — | 0.478 | 1.33 | — | Unknown | 83.9 |
| [64] | foam | 1 | 20 /25 | water | 1D interface | 99 (250-2500nm) | 38.4 | 0.36 | 1.27 | 0.132 | Yes | 87.5 |
| [65] | foam | 1 | 20 /30 | water | 2D interface | 97 ((250-1200nm)) | 36.5 | 0.39 | 1.25 | 0.1 | Yes | 85.6 |
| [66] | foam | 12 | 27 /— | salt water (3%) | 3D interface | — | 67 | — | 14.02 | — | Unknown | 82.8 |
| [67] | foam | 1 | 21 /10 | water | 2D interface | 98 (250-2500nm) | 44.2 | 0.43 | 1.28 | 0.125 | Yes | 88 |
| [68] | foam | 10 | — | water | 1D interface | 92 (300-2500nm) | 82 | — | 12.6 | — | Unknown | 89 |
| [69] | foam | 1 | 20 /60 | water | 3D interface | — | — | 0.23 | 0.83 | — | Unknown | 52.2 |
| [70] | foam | 1 | 28 /— | salt water (2.75%) | 3D interface | 90-95(295-2000nm) | — | 0.4 | 1.4 | 0.1 | Yes | 91.4 |
| [71] | foam | 1 | — | water | 2D interface | 90 (300-800nm) | 43.9 | 0.42 | 1.31 | 0.1 | Yes | 83 |
| [72] | foam | 10 | — | water | 3D interface | 91 (300-1500nm) | — | 0.95 | 11.24 | — | Unknown | 81 |
| [73] | foam | 1 | 25 /— | water | 2D interface | 90 (200-2500nm) | 40 | 0.05 | 1.282 | 0.492 | Yes | 9 |
| [74] | foam | 3.5 | 21.2 /— | water | 3D interface | 98 (200-800nm) | 75.7 | 1.8 | 7.54 | — | Yes | 135 |
| [75] | foam | 1 | 28 /41 | water | 1D interface | 96 (250-2500nm) | 38 | 0.45 | 1.27 | 0.2 | Yes | 78 |
| [76] | foam | 1 | 26 /40 | water | 3D interface | 98 (300-2500nm) | 43 | 0.45 | 1.05 | — | Unknown | 72 |
| [77] | foam | 1 | — | water | Bottom Heating | — | — | 0.33 | 0.83 | — | Unknown | — |
| [78] | foam | 10 | — | water | 3D interface | 99 (250-2500nm) | — | 3.3 | 12.2 | 0.08 | Yes | 87 |
| [79] | foam | 1 | 24 /10 | water | 3D interface | 95.2 (250-2250nm) | 47 | 0.43 | 1.16 | — | Yes | 81 |
| [80] | foam | 1 | 25 /— | water | 1D interface | 98 (250-2000nm) | 45 | 0.48 | 1.3 | 0.13 | Yes | 86.5 |
| [81] | foam | 1 | 23 /40 | water | 1D interface | 98 (400-2500nm) | 48.5 | — | 1.09 | 0.286 | Yes | 73 |
| [82] | foam | 1 | 21.5 /55 | water | 1D interface | 97 (400-2500nm) | 42 | — | 1.37 | 0.23 | Yes | 100 |
| [83] | foam | 1 | — | water | 3D interface | 97.5 (300-1200nm) | 32.7 | 0.24 | 1.11 | 0.05 | Yes | 76.3 |
| [84] | foam | 1 | — | water | 2D interface | 94 (300-2500nm) | 48 | 0.55 | 1.22 | 0.2 | Yes | 79.4 |
| [85] | foam | 1 | — | salt water (3.5%) | 2D interface | — | 42 | — | — | — | — | 56 |
| [86] | foam | 1 | 22 /— | water | 3D interface | 95 (200-2500nm) | 38 | 0.3 | — | — | Unknown | 80 |
| [87] | foam | 1 | 30 /— | water | 1D interface | 95 (250-2500nm) | 32.7 | — | — | 0.47 | — | 85 |
| [88] | foam | 1 | 25 /50 | water | 3D interface | 96 (400-1200nm) | 43 | 0.462 | — | — | Unknown | 87.3 |
| [89] | foam | 1 | 25 /— | water | 1D interface | 90-95(250-2500nm) | 42.2 | 0.502 | 1.58 | 0.224 | Yes | 84.95 |
| [90] | foam | 10 | 20 /— | water | 3D interface | 95 (300-2500nm) | 103 | — | 12.26 | — | Unknown | 89 |
| [91] | foam | 5 | — | water | 3D interface | 99(400-2500nm) | 51.5 | 1.27 | 6.6 | — | Yes | 86.2 |
| [92] | foam | 1 | 30 /60 | water | 1D interface | 97 (250-2500nm) | 50 | — | 1.45 | 0.156 | Yes | 91.3 |
| [93] | foam | 1 | 24 /40 | water | 2D interface | 95 (400-1100nm) | 48.5 | — | 1.13 | 0.11 | Yes | 78 |
| [94] | foam | 1 | 21.5 /55 | water | 1D interface | 99.4 (250-2500nm) | 37 | — | 1.45 | 0.59 | Yes | 100 |
| [95] | foam | 1 | 25 /— | water | 3D interface | 98 (250-2000nm) | 34 | 0.15 | 1 | 0.015 | Yes | 65 |
| [96] | foam | 1 | 22 /20 | water | 1D interface | 99 (250-2500nm) | 46.5 | 0.45 | 1.45 | 0.35 | Yes | 91 |
| [97] | foam | 1 | —/45 | water | 3D interface | 100 (200-2500nm) | 46 | 0.5 | 3.2 | 0.025 | Yes | 94 |
| [98] | foam | 1 | 25 /45 | water | 3D interface | 96.5(250-2500nm) | 34.3 | 0.3 | 2.15 | — | No | — |
| [99] | foam | 1 | 25 /— | water | 1D interface | 98 (400-2500nm) | 43 | 0.39 | 1.79 | 0.19 | Yes | 92 |
| [100] | foam | 1 | — | water | 3D interface | 97 (250-2000nm) | 38 | 0.44 | 2.6 | — | Unknown | 91 |
| [101] | foam | 1 | 20 /60 | salt water (20%) | 3D interface | 98 (250-2500nm) | — | — | 1.04 | — | Yes | 75 |
| [102] | foam | 1 | — | water | 3D interface | 94 (400-2500nm） | 43 | 0.463 | — | — | Unknown | 90.4 |
| [103] | bulk water | 1 | — | salt water (3.5%) | contactless | 92 (250-1500nm) | 133 | — | — | — | — | 25 |

**Note:** “E_water” is the evaporation rate of pure water under solar irradiation, “Ev_rate” is the evaporation rate of using micro/nanomaterials under solar irradiation, “Ev_dark” is the evaporation rate of using micro/nanomaterials under dark condition. “Calibrated with Ev_dark” means whether the “Ev_dark” is subtracted from “Ev_rate”. “E.F.” is the efficiency of evaporation

## References

[1] J. Carrete, W. Li, N. Mingo, S. Wang, S. Curtarolo, *Phys Rev X* **2014**, 4, 011019.

[2] A. C. Davison, D. V. Hinkley, *Bootstrap methods and their application*, Cambridge university press, **1997**.

[3] L. Breiman, *Mach Learn* **2001**, 45, 5.

[4] D. Steinberg, P. Colla, *The top ten algorithms in data mining* **2009**, 9, 179.

[5] X. Chen, M. Wang, H. Zhang, *Wiley Interdisciplinary Reviews: Data Mining and Knowledge Discovery* **2011**, 1, 55.

[6] Y. Zeng, J. Yao, B. A. Horri, K. Wang, Y. Wu, D. Li, H. Wang, *Energ Environ Sci* **2011**, 4, 4074.

[7] Y. Zeng, K. Wang, J. Yao, H. Wang, *Chem Eng Sci* **2014**, 116, 704.

[8] G. Ni, N. Miljkovic, H. Ghasemi, X. Huang, S. V. Boriskina, C.-T. Lin, J. Wang, Y. Xu, M. M. Rahman, T. Zhang, G. Chen, *Nano Energy* **2015**, 17, 290.

[9] D. Zhao, H. Duan, S. Yu, Y. Zhang, J. He, X. Quan, P. Tao, W. Shang, J. Wu, C. Song, T. Deng, *Sci Rep* **2015**, 5, 17276.

[10] S. Ishii, R. P. Sugavaneshwar, K. Chen, T. D. Dao, T. Nagao, *Opt Mater Express* **2016**, 6, 640.

[11] S. Ishii, R. P. Sugavaneshwar, T. Nagao, *The Journal of Physical Chemistry C* **2016**, 120, 2343.

[12] H. Jin, G. Lin, L. Bai, A. Zeiny, D. Wen, *Nano Energy* **2016**, 28, 397.

[13] X. Wang, Y. He, G. Cheng, L. Shi, X. Liu, J. Zhu, *Energ Convers Manage* **2016**, 130, 176.

[14] X. Wang, G. Ou, N. Wang, H. Wu, *ACS Appl Mater Interfaces* **2016**, 8, 9194.

[15] M. S. Zielinski, J. W. Choi, T. La Grange, M. Modestino, S. M. Hashemi, Y. Pu, S. Birkhold, J. A. Hubbell, D. Psaltis, *Nano Lett* **2016**, 16, 2159.

[16] M. Amjad, G. Raza, Y. Xin, S. Pervaiz, J. Xu, X. Du, D. Wen, *Appl Energ* **2017**, 206, 393.

[17] Y. Fu, T. Mei, G. Wang, A. Guo, G. Dai, S. Wang, J. Wang, J. Li, X. Wang, *Appl Therm Eng* **2017**, 114, 961.

[18] H. Li, Y. He, Z. Liu, Y. Huang, B. Jiang, *Appl Therm Eng* **2017**, 121, 617.

[19] L. Shi, Y. He, Y. Huang, B. Jiang, *Energ Convers Manage* **2017**, 149, 401.

[20] X. Wang, Y. He, X. Liu, L. Shi, J. Zhu, *Sol Energy* **2017**, 157, 35.

[21] Z. Wang, Y. Liu, P. Tao, Q. Shen, N. Yi, F. Zhang, Q. Liu, C. Song, D. Zhang, W. Shang, T. Deng, *Small* **2014**, 10, 3234.

[22] K. Bae, G. Kang, S. K. Cho, W. Park, K. Kim, W. J. Padilla, *Nat Commun* **2015**, 6, 10103.

[23] Y. Ito, Y. Tanabe, J. Han, T. Fujita, K. Tanigaki, M. Chen, *Adv Mater* **2015**, 27, 4302.

[24] Y. Liu, S. Yu, R. Feng, A. Bernard, Y. Liu, Y. Zhang, H. Duan, W. Shang, P. Tao, C. Song, T. Deng, *Adv Mater* **2015**, 27, 2768.

[25] L. Zhang, B. Tang, J. Wu, R. Li, P. Wang, *Adv Mater* **2015**, 27, 4889.

[26] P. Fan, H. Wu, M. Zhong, H. Zhang, B. Bai, G. Jin, *Nanoscale* **2016**, 8, 14617.

[27] Z. Hua, B. Li, L. Li, X. Yin, K. Chen, W. Wang, *The Journal of Physical Chemistry C* **2016**, 121, 60.

[28] Y. Liu, J. Lou, M. Ni, C. Song, J. Wu, N. P. Dasgupta, P. Tao, W. Shang, T. Deng, *ACS Appl Mater Interfaces* **2016**, 8, 772.

[29] Y. Wang, L. Zhang, P. Wang, *Acs Sustain Chem Eng* **2016**, 4, 1223.

[30] C. Zhang, C. Yan, Z. Xue, W. Yu, Y. Xie, T. Wang, *Small* **2016**, 12, 5320.

[31] L. Zhou, Y. Tan, D. Ji, B. Zhu, P. Zhang, J. Xu, Q. Gan, Z. Yu, J. Zhu, *Sci. Adv.* **2016**, 2, e1501227

[32] L. Zhou, Y. Tan, J. Wang, W. Xu, Y. Yuan, W. Cai, S. Zhu, J. Zhu, *Nat Photonics* **2016**, 10, 393.

[33] G. Zhu, J. Xu, W. Zhao, F. Huang, *ACS Appl Mater Interfaces* **2016**, 8, 31716.

[34] C. Chen, Y. Li, J. Song, Z. Yang, Y. Kuang, E. Hitz, C. Jia, A. Gong, F. Jiang, J. Y. Zhu, B. Yang, J. Xie, L. Hu, *Adv Mater* **2017**, 29.

[35] D. Ding, W. Huang, C. Song, M. Yan, C. Guo, S. Liu, *Chem Commun (Camb)* **2017**, 53, 6744.

[36] X. Gao, H. Ren, J. Zhou, R. Du, C. Yin, R. Liu, H. Peng, L. Tong, Z. Liu, J. Zhang, *Chem Mater* **2017**, 29, 5777.

[37] V. Kashyap, A. Al-Bayati, S. M. Sajadi, P. Irajizad, S. H. Wang, H. Ghasemi, *J Mater Chem A* **2017**, 5, 15227.

[38] C. Liu, J. Huang, C.-E. Hsiung, Y. Tian, J. Wang, Y. Han, A. Fratalocchi, *Advanced Sustainable Systems* **2017**, 1, 1600013.

[39] G. Wang, Y. Fu, X. Ma, W. Pi, D. Liu, X. Wang, *Carbon* **2017**, 114, 117.

[40] J. Wang, Z. Liu, X. Dong, C.-E. Hsiung, Y. Zhu, L. Liu, Y. Han, *J Mater Chem A* **2017**, 5, 6860.

[41] X. Wang, Y. He, X. Liu, G. Cheng, J. Zhu, *Appl Energ* **2017**, 195, 414.

[42] J. Yang, Y. Pang, W. Huang, S. K. Shaw, J. Schiffbauer, M. A. Pillers, X. Mu, S. Luo, T. Zhang, Y. Huang, G. Li, S. Ptasinska, M. Lieberman, T. Luo, *Acs Nano* **2017**, 11, 5510.

[43] L. Yi, S. Ci, S. Luo, P. Shao, Y. Hou, Z. Wen, *Nano Energy* **2017**, 41, 600.

[44] Z. Yin, H. Wang, M. Jian, Y. Li, K. Xia, M. Zhang, C. Wang, Q. Wang, M. Ma, Q. S. Zheng, Y. Zhang, *ACS Appl Mater Interfaces* **2017**, 9, 28596.

[45] P. Zhang, J. Li, L. Lv, Y. Zhao, L. Qu, *Acs Nano* **2017**, 11, 5087.

[46] T. F. Chala, C. M. Wu, M. H. Chou, Z. L. Guo, *ACS Appl Mater Interfaces* **2018**, 10, 28955.

[47] M. Chen, Y. Wu, W. Song, Y. Mo, X. Lin, Q. He, B. Guo, *Nanoscale* **2018**, 10, 6186.

[48] F. Tao, Y. Zhang, K. Yin, S. Cao, X. Chang, Y. Lei, D. S. Wang, R. Fan, L. Dong, Y. Yin, X. Chen, *ACS Appl Mater Interfaces* **2018**, DOI: 10.1021/acsami.8b11786.

[49] W. Xu, X. Hu, S. Zhuang, Y. Wang, X. Li, L. Zhou, S. Zhu, J. Zhu, *Adv Energy Mater* **2018**, 8, 1702884.

[50] X. Yang, Y. Yang, L. Fu, M. Zou, Z. Li, A. Cao, Q. Yuan, *Adv Funct Mater* **2018**, 28, 1704505.

[51] H. Ghasemi, G. Ni, A. M. Marconnet, J. Loomis, S. Yerci, N. Miljkovic, G. Chen, *Nat Commun* **2014**, 5, 4449.

[52] F. M. Canbazoglu, B. Fan, A. Kargar, K. Vemuri, P. R. Bandaru, *Aip Adv* **2016**, 6, 085218.

[53] Q. Jiang, L. Tian, K. K. Liu, S. Tadepalli, R. Raliya, P. Biswas, R. R. Naik, S. Singamaneni, *Adv Mater* **2016**, 28, 9400.

[54] X. Li, W. Xu, M. Tang, L. Zhou, B. Zhu, S. Zhu, J. Zhu, *Proc Natl Acad Sci U S A* **2016**, 113, 13953.

[55] G. Ni, G. Li, Svetlana V. Boriskina, H. Li, W. Yang, T. Zhang, G. Chen, *Nature Energy* **2016**, 1, 16126.

[56] S. M. Sajadi, N. Farokhnia, P. Irajizad, M. Hasnain, H. Ghasemi, *J Mater Chem A* **2016**, 4, 4700.
[57] L. Tian, J. Luan, K. K. Liu, Q. Jiang, S. Tadepalli, M. K. Gupta, R. R. Naik, S. Singamaneni, *Nano Lett* **2016**, 16, 609.
[58] Y. Fu, G. Wang, T. Mei, J. Li, J. Wang, X. Wang, *Acs Sustain Chem Eng* **2017**, 5, 4665.
[59] X. Hu, W. Xu, L. Zhou, Y. Tan, Y. Wang, S. Zhu, J. Zhu, *Adv Mater* **2017**, 29.
[60] C. Jia, Y. Li, Z. Yang, G. Chen, Y. Yao, F. Jiang, Y. Kuang, G. Pastel, H. Xie, B. Yang, S. Das, L. Hu, *Joule* **2017**, 1, 588.
[61] Q. Jiang, H. Gholami Derami, D. Ghim, S. Cao, Y.-S. Jun, S. Singamaneni, *J Mater Chem A* **2017**, 5, 18397.
[62] M. Kaur, S. Ishii, S. L. Shinde, T. Nagao, *Acs Sustain Chem Eng* **2017**, 5, 8523.
[63] R. Li, L. Zhang, L. Shi, P. Wang, *Acs Nano* **2017**, 11, 3752.
[64] Y. Li, T. Gao, Z. Yang, C. Chen, Y. Kuang, J. Song, C. Jia, E. M. Hitz, B. Yang, L. Hu, *Nano Energy* **2017**, 41, 201.
[65] Y. Li, T. Gao, Z. Yang, C. Chen, W. Luo, J. Song, E. Hitz, C. Jia, Y. Zhou, B. Liu, B. Yang, L. Hu, *Adv Mater* **2017**, 29.
[66] K. K. Liu, Q. Jiang, S. Tadepalli, R. Raliya, P. Biswas, R. R. Naik, S. Singamaneni, *ACS Appl Mater Interfaces* **2017**, 9, 7675.
[67] Z. Liu, H. Song, D. Ji, C. Li, A. Cheney, Y. Liu, N. Zhang, X. Zeng, B. Chen, J. Gao, Y. Li, X. Liu, D. Aga, S. Jiang, Z. Yu, Q. Gan, *Global Chall* **2017**, 1, 1600003.
[68] Z. Liu, Z. Yang, X. Huang, C. Xuan, J. Xie, H. Fu, Q. Wu, J. Zhang, X. Zhou, Y. Liu, *J Mater Chem A* **2017**, 5, 20044.
[69] S. Ma, C. P. Chiu, Y. Zhu, C. Y. Tang, H. Long, W. Qarony, X. Zhao, X. Zhang, W. H. Lo, Y. H. Tsang, *Appl Energ* **2017**, 206, 63.
[70] H. Ren, M. Tang, B. Guan, K. Wang, J. Yang, F. Wang, M. Wang, J. Shan, Z. Chen, D. Wei, H. Peng, Z. Liu, *Adv Mater* **2017**, 29, 1702590.
[71] L. Shi, Y. Wang, L. Zhang, P. Wang, *J Mater Chem A* **2017**, 5, 16212.
[72] G. Wang, Y. Fu, A. Guo, T. Mei, J. Wang, J. Li, X. Wang, *Chem Mater* **2017**, 29, 5629.
[73] Z. Wang, Q. Ye, X. Liang, J. Xu, C. Chang, C. Song, W. Shang, J. Wu, P. Tao, T. Deng, *J Mater Chem A* **2017**, 5, 16359.
[74] X. Wu, G. Y. Chen, W. Zhang, X. Liu, H. Xu, *Advanced Sustainable Systems* **2017**, 1, 1700046.
[75] N. Xu, X. Hu, W. Xu, X. Li, L. Zhou, S. Zhu, J. Zhu, *Adv Mater* **2017**, 29.
[76] G. Xue, K. Liu, Q. Chen, P. Yang, J. Li, T. Ding, J. Duan, B. Qi, J. Zhou, *ACS Appl Mater Interfaces* **2017**, 9, 15052.
[77] J. D. Yao, Z. Q. Zheng, G. W. Yang, *Nanoscale* **2017**, 9, 16396.
[78] M. Zhu, Y. Li, G. Chen, F. Jiang, Z. Yang, X. Luo, Y. Wang, S. D. Lacey, J. Dai, C. Wang, C. Jia, J. Wan, Y. Yao, A. Gong, B. Yang, Z. Yu, S. Das, L. Hu, *Adv Mater* **2017**, 29.
[79] Q. Chen, Z. Pei, Y. Xu, Z. Li, Y. Yang, Y. Wei, Y. Ji, *Chem Sci* **2018**, 9, 623.
[80] J. Fang, J. Liu, J. Gu, Q. Liu, W. Zhang, H. Su, D. Zhang, *Chem Mater* **2018**, 30, 6217.
[81] X. Gao, H. Lan, S. Li, X. Lu, M. Zeng, X. Gao, Q. Wang, G. Zhou, J.-M. Liu, M. J. Naughton, K. Kempa, J. Gao, *Global Challenges* **2018**, DOI: 10.1002/gch2.2018000351800035.
[82] S. Hong, Y. Shi, R. Li, C. Zhang, Y. Jin, P. Wang, *ACS Appl Mater Interfaces* **2018**, 10,

28517.
[83] F. Jiang, H. Liu, Y. Li, Y. Kuang, X. Xu, C. Chen, H. Huang, C. Jia, X. Zhao, E. Hitz, Y. Zhou, R. Yang, L. Cui, L. Hu, *ACS Appl Mater Interfaces* **2018**, 10, 1104.
[84] H. Li, Y. He, Y. Hu, X. Wang, *ACS Appl Mater Interfaces* **2018**, 10, 9362.
[85] G. Ni, S. H. Zandavi, S. M. Javid, S. V. Boriskina, T. A. Cooper, G. Chen, *Energ Environ Sci* **2018**, 11, 1510.
[86] T. Li, H. Liu, X. Zhao, G. Chen, J. Dai, G. Pastel, C. Jia, C. Chen, E. Hitz, D. Siddhartha, R. Yang, L. Hu, *Adv Funct Mater* **2018**, 28, 1707134.
[87] X. Li, R. Lin, G. Ni, N. Xu, X. Hu, B. Zhu, G. Lv, J. Li, S. Zhu, J. Zhu, *Natl Sci Rev* **2018**, 5, 70.
[88] X. Lin, J. Chen, Z. Yuan, M. Yang, G. Chen, D. Yu, M. Zhang, W. Hong, X. Chen, *J Mater Chem A* **2018**, 6, 4642.
[89] F. Liu, B. Zhao, W. Wu, H. Yang, Y. Ning, Y. Lai, R. Bradley, *Adv Funct Mater* **2018**, 28, 1803266.
[90] H. Liu, C. Chen, G. Chen, Y. Kuang, X. Zhao, J. Song, C. Jia, X. Xu, E. Hitz, H. Xie, S. Wang, F. Jiang, T. Li, Y. Li, A. Gong, R. Yang, S. Das, L. Hu, *Adv Energy Mater* **2018**, 8, 1701616.
[91] H. Liu, C. Chen, H. Wen, R. Guo, N. A. Williams, B. Wang, F. Chen, L. Hu, *J Mater Chem A* **2018**, 6, 18839.
[92] P.-F. Liu, L. Miao, Z. Deng, J. Zhou, H. Su, L. Sun, S. Tanemura, W. Cao, F. Jiang, L.-D. Zhao, *Materials Today Energy* **2018**, 8, 166.
[93] G. Peng, H. Ding, S. W. Sharshir, X. Li, H. Liu, D. Ma, L. Wu, J. Zang, H. Liu, W. Yu, H. Xie, N. Yang, *Appl Therm Eng* **2018**, 143, 1079.
[94] Y. Shi, R. Li, Y. Jin, S. Zhuo, L. Shi, J. Chang, S. Hong, K.-C. Ng, P. Wang, *Joule* **2018**, 2, 1171.
[95] C. H. Xiao Luo, Shang Liu, Jinxin Zhong, *Int. J. Energy Res.* **2018**, 42, 4830.
[96] P. Zhang, Q. Liao, T. Zhang, H. Cheng, Y. Huang, C. Yang, C. Li, L. Jiang, L. Qu, *Nano Energy* **2018**, 46, 415.
[97] F. Zhao, X. Zhou, Y. Shi, X. Qian, M. Alexander, X. Zhao, S. Mendez, R. Yang, L. Qu, G. Yu, *Nat Nanotechnol* **2018**, 13, 489.
[98] Z. Deng, L. Miao, P.-F. Liu, J. Zhou, P. Wang, Y. Gu, X. Wang, H. Cai, L. Sun, S. Tanemura, *Nano Energy* **2019**, 55, 368.
[99] F. Gong, H. Li, W. Wang, J. Huang, D. Xia, J. Liao, M. Wu, D. V. Papavassiliou, *Nano Energy* **2019**, 58, 322.
[100] Y. Guo, F. Zhao, X. Zhou, Z. Chen, G. Yu, *Nano Lett* **2019**, 19, 2530.
[101] Y. Kuang, C. Chen, S. He, E. M. Hitz, Y. Wang, W. Gan, R. Mi, L. Hu, *Adv Mater* **2019**, DOI: 10.1002/adma.201900498e1900498.
[102] C. Li, D. Jiang, B. Huo, M. Ding, C. Huang, D. Jia, H. Li, C.-Y. Liu, J. Liu, *Nano Energy* **2019**, 60, 841.
[103] T. A. Cooper, S. H. Zandavi, G. W. Ni, Y. Tsurimaki, Y. Huang, S. V. Boriskina, G. Chen, *Nat Commun* **2018**, 9, 5086.